\documentclass[12pt,preprint]{aastex}
\def \msun          {\hbox{M$_\odot$}}
\def \Mo            {\hbox{M$_\odot$}}
\def \kms          {km s$^{-1}$}

\newcommand\pzz  {\phantom {00}}
\newcommand\pz   {\phantom {0}}

\begin{document}

\title{The Mass Ratio Distribution in Main-Sequence Spectroscopic Binaries
Measured by IR Spectroscopy}

\author{T. Mazeh\altaffilmark{1}, M. Simon\altaffilmark{2}, L. 
Prato\altaffilmark{3}, B. Markus\altaffilmark{1}, and S. 
Zucker\altaffilmark{4}}

\altaffiltext{1}{Department of Physics and Astronomy, Tel Aviv University,
Tel Aviv 69978, Israel; mazeh@wise.tau.ac.il, barak@wise.tau.ac.il}

\altaffiltext{2}{Department of Physics and Astronomy,
SUNY$-$SB, Stony Brook, NY 11794-3800; michal.simon@sunysb.edu}

\altaffiltext{3}{Department of Physics and Astronomy,
University of California, Los Angeles, CA, 90095-1562; lprato@astro.ucla.edu}

\altaffiltext{4}{Department of Geophysics and Planetary Sciences, Tel Aviv
University, Tel Aviv, Israel; shay@wise.tau.ac.il}

\begin{abstract}
  
We report infrared spectroscopic observations of a large, well-defined
sample of main-sequence, single-lined spectroscopic binaries in order
to detect the secondaries and derive the mass ratio distribution of
short-period binaries. The sample consists of 51 Galactic disk
spectroscopic binaries found in the Carney \& Latham
high-proper-motion survey, with primary masses in the range of
0.6--0.85 \msun.  Our infrared observations detect the secondaries in
32 systems, two of which have mass ratios, $q=M_2/M_1$, as low as
$\sim0.20$.  Together with 11 systems previously identified as
double-lined binaries by visible light spectroscopy, we have a
complete sample of 62 binaries, out of which 43 are double-lined.  The
mass ratio distribution is approximately constant over the range
$q=1.0$ to 0.3. The distribution appears to rise at lower $q$ values,
but the uncertainties are sufficiently large that we cannot rule out a
distribution that remains constant.  The mass distribution derived for
the secondaries in our sample, and that of the extra-solar planets,
apparently represent two distinct populations.

\end{abstract}


\section{Introduction}

The mass ratio distribution of binaries provides one of the few
diagnostics for testing models of binary formation (for recent
reviews, see Clarke 2001; Tohline 2002; Halbwachs et al.\ 2003).
However, the use of the mass ratio distribution in spectroscopic
binaries (SBs) has been limited because, in most studies, only the
primary spectrum is detected.  Without the secondary radial
velocities, the mass ratio of a spectroscopic binary cannot be
measured.  Detection of the secondary spectrum is difficult because
most radial velocity observations are made in visible light, where the
luminosity of stars less massive than $\sim 1$ \msun\ is a strong
function of the mass (e.g., Carney et al.\ 1994, hereafter CL94). As a
result, the measured secondary/primary mass ratios, $q=M_2/M_1$, are
clustered close to 1.  To overcome this bias astronomers have applied
powerful statistical techniques to samples of single and double-lined
spectroscopic binaries (SB1s and SB2s), in order to derive the mass
ratio distribution down to mass ratios as low as $\sim 0.1$ (e.g.,
Halbwachs 1987; Heacox 1995). However, because of the missing
information on the mass ratios of the binaries, the resolution ability
of the statistical techniques is limited, and they can only derive the
gross features of the distribution. This might be the reason why the
results of such analyses still differ and a consensus on the
underlying distribution has not been reached.

Goldberg, Mazeh \& Latham (2003, hereafter G03) studied 129 binaries
identified in the Carney \& Latham high-proper-motion sample (CL94),
which consisted of a large fraction of halo stars.  Their sample of
binaries included 25 SB2s.  G03 found a rise in the distribution as
the mass ratio decreases to $q\sim 0.2$, a drop at $q< 0.2$, and a
small peak at $q\sim 0.8$.  Halbwachs et al.\ (2003) considered a
well-defined sample of 27 binaries within the nearby F7 to K stars,
together with 25 binaries found in the open clusters of Pleiades and
Praesepe. Their sample of 52 binaries consisted of 15 SB2s, for which
the mass ratios were directly derived, another 9 binaries with mass
ratios estimated from astrometric data (Udry et al.\ 2003), and an
additional 7 cluster binaries with mass ratios estimated from their
photometric properties.  Halbwachs et al.\ found a mass ratio
distribution with two peaks: a broad, shallow, peak between $q\sim
0.2$ and $\sim 0.7$, and a sharp, high, peak at $q > 0.8$. The latter
appears only in the distribution of the short period binaries, with
periods shorter than 50 days.
 
Evidently, even the most recent studies do not agree on the true mass
ratio distribution.  It is therefore desirable to derive the mass
ratio distribution with minimal need for statistical tools.  To
accomplish this we need a large, well defined sample of binaries
dominated by SB2s with mass ratios measured over as large a range as
possible.  Our approach to the detection of main-sequence and
pre$-$main-sequence (PMS) SB2s is to observe with high resolution,
infrared (IR) spectroscopy SB1s that were previously identified and
measured with visible light (Mazeh et al.\ 2002; Prato et al.\
2002a,b).  Because the secondaries in SB1s are cool, the secondary to
primary light ratio is greater in the IR than in visible band,
favoring their detection in the IR.  One result of this approach is
the smallest mass ratio ever measured dynamically in a PMS SB,
$q=0.18\pm0.01$, in the NTTS 160905$-$1859 system (Prato et al.\
2002a).

To compile as complete and large as possible a sample for study, we
chose to use a subsample of the SBs in the Carney \& Latham study of
1464 high-proper-motion, nearby, late G and early K, main-sequence
stars (CL94).  Systematic, long term radial velocity monitoring of
these stars in visible light has identified 171 SB1s and 34 SB2s
(Latham et al.\ 2002, hereafter L02; Goldberg et al.\ 2002, hereafter
G02).  Out of these, we focused on 51 SB1s, turning 32 systems into
SB2s.  The composition of our sample is described in \S 2, the
observations in \S 3, the analysis in \S 4, and the measured and
derived mass ratio distribution in \S 5.  We compare our results with
previous studies in \S 6, discuss our results in \S 7 and conclude the
discussion in \S 8.

\section{The Sample}

We selected SB1s from L02 with a well defined range of metallicities
and distances, as derived by CL94 using photometry, color indices in
the visible and IR, and low S/N spectroscopy.  To exclude the halo
stars from our sample and to include stars with relatively strong
spectral lines we chose only stars identified by G03 as Galactic disk
stars, with metallicities $[Fe/H] \stackrel{>}\sim -1$.  To ensure
high S/N ratio in the observations, and to choose a volume limited
sample, we excluded all stars whose distances were estimated by CL94
to be larger than 90 pc.  The SB sample identified by L02 and G02 was
not complete for binaries with long periods (see discussion in G03),
so we chose only binaries with periods shorter than 3000 days.

Table 1 lists the 52 SB1s observed.  The system $V$-band magnitude,
metallicity, estimated distance, and binary period, taken from L02 and
CL94, appear in columns (4)--(7).  Our sample is characterized by
$V<11.46$ mag (which corresponds to $H\stackrel{<}\sim 9.6$ mag for these 
stars) and $[Fe/H] > -1.09$.   Column (8) of Table 1 lists the mass
estimates of the primaries, as derived by CL94.

Analysis of a single-lined orbit yields the mass function,

$$f(M) = {(M_2 ~sin~i)^3\over(M_1+M_2)^2} = {q^3\over(1+q)^2}M_1sin^3i $$

\noindent where $i$ is the orbital inclination.  Using the estimate of
the primary mass and the limit of an edge on orbit, $i=\pi/2$, $f(M)$
determines the minimum mass ratio $q_{min}$.  We used the mass
functions reported by L02 to calculate the values of $q_{min}$ for the
SB1s in our sample and listed these in column (8) of Table~1.  The
uncertainties include that of the mass function and our estimate of
$\pm 0.1$ \msun\ for the uncertainty of $M_{1,est}$.  During the
analysis we found that one system in Table~1, G17-22, has a minimum
mass ratio substantially larger than unity, $1.36 \pm 0.13$. This
probably indicates a white dwarf secondary that was originally the
primary star. If this is true, the mass ratio of the system now
differs from its original value.  We therefore excluded this system
from further analysis.

For completeness we included in our analysis the SB2s already detected
by G02 which have the same limits on distance, metallicity, and period
as the SB1s in Table~1.  Table 2 lists these 11 binaries and is
organized in the same way as Table 1, except that the last column
lists the mass ratio measured by G02. Our study of the mass ratio
distribution of nearby main-sequence SBs is thus based on 62 systems,
the 11 SB2s in Table 2 with measured $q$'s, all close to 1, and the 51
SB1s in Table 1 (excluding G17-22). Figure 1 shows the distribution of
$M_{1,est}$ in the entire sample of 62 SBs.  The range of the masses
is quite narrow, 0.6--0.85 \Mo, with a median value of $0.72$ \Mo.

Our sample definition relies on CL94's distances rather than {\it
HIPPARCOS} measurements, because only the former are available for all
the systems in our sample, and the latter have large uncertainties at
distances greater than $\sim 60$pc. Photometric distances might be
biased towards smaller distances for binaries with high mass ratios,
as the secondaries make such binaries brighter than single stars. Such
bias could have introduced the \"Opik effect (\"Opik 1926; Branch
1976) into our sample of binaries. To look for such an effect we
considered the ratio of the Hipparcos distance over the CL
distance for each of the binaries in our sample. If the \"Opik was
present in our sample we would expect the ratio to be larger for
binaries with mass ratio close to unity.
To investigate this possibility, we divided our sample into two
subsamples, with $q_{min}$ above and below $0.85$.  The average and
r.m.s of the ratios of the high and low mass ratio samples were $1.12
\pm 0.26$ and $1.29 \pm 0.39$, respectively. The difference between
the two figures is insignificant, and certainly not in the direction
expected for the \"Opik effect. We conclude that there is no evidence
that the \"Opik effect has contaminated our sample.

\section{Observations and Data Reduction}

Observations of all stars in Table 1 were made with the NIRSPEC
spectrometer at the Keck II telescope (McLean et al.\ 1998, 2000),
following the procedures described by Prato et al.\ (2002a,b). 
The spectrometer was centered on 1.55 $\mu$m
in order 49.  Exposure times were long enough to obtain spectra with
S/N $\ga100$ on each target, typically between 5 and 10 minutes.  The
observations in June, 2000, and January, 2001, were obtained using
adaptive optics (AO) and achieved resolution R $\sim30,000$.
Observations in February, May, and June, 2001, were made without the AO system
and have resolution R $\sim24,000$.  A few more spectra for some of
the systems were obtained when time permitted in subsequent observing
runs; these later observations were all in the non-AO mode.  The
modified Julian dates of observation for each binary in the sample are
listed in Table 3, column (2).  The spectra were extracted and
wavelength calibrated using the REDSPEC software package\footnote{See:
http://www2.keck.hawaii.edu/inst/nirspec/redspec/index.html}.  We used
only order 49 for the results presented here because it is almost
completely free of terrestrial absorption lines.

As in our previous studies (Mazeh et al.\ 2002; Prato et al.\ 2002a,b),
we used the TODCOR two-dimensional correlation algorithm (Zucker
\& Mazeh 1994) and our library of stellar templates to analyze the
target spectra for the velocities of the primary and secondary stars.
Table~3 lists the derived velocities of the primary, $v_1$, and the
secondary, $v_2$, and their uncertainties, as yielded by TODCOR. The
uncertainties do not include the errors induced by the scatter of the
radial velocities assigned to the templates, which we estimate to be
of the order of 1 km s$^{-1}$. The latter were added when
we used these velocities for the orbital solution.

The primary velocity was detectable in every observation of each
target, whereas not all spectra yielded a reliable measurement of the
secondary velocities. This is because the correlation peak of the
secondary was not always sufficiently prominent because of
several possible factors including the secondary to primary light
ratio, the inevitable differences between the spectra of the SB
components and their corresponding templates, small velocity
differences between the components, and the S/N ratio of
the observed spectra.  The last two factors may vary between one
observation and another, which explains why a reliable secondary
velocity may be obtained from one spectrum of an object and not from
another.  For our analysis we chose only spectra that yielded
correlation functions for which we judged the secondary peak to be
well determined.  An example of the derived correlation is given in
Figure 2, which shows two aspects of the two-dimensional correlation
function of one spectrum of G121-75, for which we estimate the flux
ratio to be $0.08 \pm 0.01$. The relatively small flux ratio caused
the increase of the correlation function at the secondary velocity to
be minute, from $\sim 0.935$ to $\sim 0.945$, as can be seen from the
lower panel of the figure. Nevertheless, the peak is well above the
noise.

Our criteria led us to accept only 58 secondary velocities out of 112
observations.  The primary and secondary velocities, and their $1
\sigma$ uncertainties are listed in Table~3, columns (3)--(6). Out of
the 51 systems in the IR sample, we have derived the secondary radial
velocities for 32 systems.  For convenience, columns (2) and (3) of
Table 4 list the numbers of times that a target was observed and the
secondary detected.

\section{Double-Lined Orbital Analysis}

In all the 32 binaries for which we measured the secondary velocity,
we derived new SB2 orbital solutions using our measured values of
$v_1$ and $v_2$ together with the primary velocities of L02.  All the
newly derived elements were consistent with the SB1
elements of L02.  The mass ratios of each binary, $q_{orb}$, were
derived from the ratio of the primary and secondary semi-amplitudes,
$K_1$ and $K_2$, and are listed in column (4) of Table 4.  $K_1$ was
determined by combining the many radial velocities of L02 with our few
additional measurements, and is therefore well established.  On the
other hand, $K_2$ was determined only by our few secondary velocity
measurements, and is therefore less robust.  We have more than one
radial velocity measurement of the secondary for only 14 systems.
Examples of the orbital solutions for 4 binaries are plotted in Figure
3. All $v_2$ measurements in these 14 systems are within 1$\sigma$ of the
best fit orbital solution.

For 18 of the systems we have only one $v_2$ measurement, 4 of which
have  $q_{orb}$ values of low statistical significance,
$q_{orb}/\sigma < 2$.  Two tests give us confidence that the set of
$q_{orb}$'s we have derived are reliable.  Clearly, we expect $q_{min}
< q_{orb}$.  Figure 4 plots $q_{min}$ as a function of $q_{orb}$, 
showing that the derived values of $q_{orb}$ are either greater than
their corresponding value of $q_{min}$ or within 2$\sigma$ of the
difference.  Using the derived $q_{orb}$ values with the corresponding
estimated masses, we can also derive $\sin i$.  In Figure 5 we plot
the distribution of the derived $\sin i$, together with the expected
distribution for a sample of randomly oriented orbits in space. The
two are very similar, again supporting our working assumption that the
derived $q_{orb}$ are correct.

TODCOR analysis also provides the flux ratio, $\alpha$, at which the
combination of shifted templates for the secondary and primary stars
yields the maximum correlation.  The last column of Table~4 lists our
best estimate for $\alpha$ at $1.55\mu$m, averaged over several
measurements when they were available. The derived ratios cannot be
exact because of the unavoidable mismatch between the actual effective
temperatures and metallicities of the primary and secondary stars and
those of the templates used in the TODCOR analysis.  Nevertheless, the
derived values give a rough estimate of the true values.  To show that
this is the case, Figure~6 plots the estimated flux ratio as a function
of the measured mass ratio.  The solid line shows the light ratio in
the photometric $H$-band for Baraffe et al.'s (1998) models of low-mass
main-sequence stars at age 5 Byr with [Fe/H]=0, referenced to primary
mass of 0.7~\msun.  Our measured values are roughly consistent with the
theoretical values.  The large scatter is probably attributable to the
approximate nature of the measured $\alpha$, the range of
metallicities and ages of the targets, and to the fact that the model
values integrate stellar spectra over the $\sim0.28 \mu$m width of the
$H$-band filter while our spectra in NIRSPEC order 49 span only $\sim
0.022\mu$m.

Figure 6 shows that detection of secondaries by IR spectroscopy
extends down to $\alpha\sim 0.04$.  Below this light ratio, our spectra in
NIRSPEC order 49 do not yield reliable secondary velocities through
the analysis described here.  As we discuss in the next section, this
is associated directly with the minimum mass ratio to which our
observations are sensitive.

\section{The Mass Ratio Distribution}

Our new measurements of the mass ratio extend down to $q=0.20 \pm
0.04$ and $0.18 \pm 0.03$ for G14-54 and G255-45, respectively (Table
4). To our knowledge, these are the lowest mass ratios yet measured
dynamically in a detached main-sequence binary in which the primary is
a G or K spectral type star.  They are matched only by our work on
PMS stars (Prato et al.\ 2002a). Together with the 11 SB2s in Table 2,
dynamically derived mass ratios are now available for 43 of the 62
binaries in our sample.

What is the nature of the binaries in which IR spectroscopy did not
detect the secondary?  There seem to be no significant differences in
the distributions of brightness, period, or metallicity (Figure
7a,b,c) of the binaries with and without IR detections of the
secondary. Applying a two-sample Kolmogorov-Smirnov (K--S) test to these
three distribution pairs yields significance values of $0.47$,
$0.45$ and $0.15$, respectively. Figure 7d suggests, however, that the 
systems with secondaries undetected in the IR  have systematically 
lower $q_{min}$'s than those with detected secondaries. The K--S test
shows that difference in the two distributions is very significant; the 
probability that they are drawn from the same distribution is 
only $5.6 \times 10^{-4}$. This suggests that systems with low $q_{min}$, 
and hence small $M_2$, are undetected probably because their 
secondary/primary light ratio falls below our sensitivity limit.

The lowest panel of Figure 7d displays our detection limit more
directly.  For each bin of width $\Delta q_{min}=0.10$, the abscissa
plots the ratio of the number of binaries with IR detected secondaries
to the number of binaries having $q_{min}$ in that bin.  The figure
indicates that our observations detected all systems with $q_{min}>
0.5$, and have monotonic decreasing sensitivity below that value.
Taken together, the distribution of non-detections in the middle
panel, and the sensitivity plotted in the lower panel, indicate that
most of the undetected secondaries in binaries with $q_{min}$ below
$0.5$ probably have actual mass ratios that are smaller than 0.5.  If
their $q$'s were larger, we probably would have detected them.

The top and middle panels of Figure 8 show the distribution of mass
ratios of the SB2s as detected in visible light by G02 and in the IR,
as described here.  To allow for the uncertainties of the mass ratios
we distributed each binary as a Gaussian centered on its measured $q$
value, with a width equal to its estimated uncertainty.  The Gaussians
were truncated at $q_{min}-\sigma_q$ and 1, where $\sigma_q$ is the
uncertainty of $q_{min}$, and were normalized accordingly.  This
smoothing procedure mitigates the influence of the few objects in our
sample with measured $q_{orb}$ values with low statistical
significance. The SB2s detected in the visible band are clustered
around $q\sim 0.85$, with a narrow spread of $\sim 0.1$, whereas the
SB2s detected in the IR are centered at $q=0.55$, with a very large
spread.

The lowest panel of Figure~8 shows our best estimate of the
distribution of the 19 SB1s that remain in our sample, derived by the
algorithm of Mazeh \& Goldberg (1992), assuming random orientations of
the binaries. We did not use the fact that the IR observations could
reliably detect a binary with mass ratio larger than $q\sim 0.5$. It
is satisfying that the derived distribution is consistent with this
limit. It is interesting to compare this estimated distribution with
that of the $q_{min}$'s of the same sample (middle panel of
Figure~7d).  The two are not very different. This is probably because
all the systems with small inclinations, and therefore with high mass
ratios, were turned into SB2s by our technique, and subsequently were
moved to the middle panel of the figure. Thus, the binaries with
undetected secondaries have on the average only slightly higher mass
ratios than their minimum values.  The top panel of Figure~9 shows the
sum of the three distributions of Figure~8.  This is an estimate of
the directly measured mass ratio distribution.

To complete the analysis, we must correct the distribution at the top
of Figure~9 for the binaries undetected by L02 and G02.  Binaries
could have escaped detection because their primary velocity
semi-amplitude, $K_1$, fell below a certain threshold.  For a given
sample of binaries with random orientations, this observational
selection effect primarily affects binaries with small mass ratios,
and so evidently biases the measured distribution (Mazeh, Latham, \&
Stefanik 1996). The correction Mazeh et al.\ suggest relies on the
assumption that the mass ratio and period distributions are
independent, and strongly depends on the period distribution of the
sample, because the effect is much stronger for longer periods.
However, the observed period distribution of the sample is subject to
the same effect, and therefore cannot be used to estimate the true
period distribution (e.g., Halbwachs et al.\ 2003). Thus, although the
selection effect is certainly there, the algorithm to correct for it
includes inherently large uncertainties.

We corrected for the undetected binaries with the algorithm presented
by Mazeh, Latham, \& Stefanik (1996), following G03 by assuming a
period distribution which is linear in log $P$ with the same
parameters.  We further assumed that the $K_1$ detection threshold is
2.5 km s$^{-1}$.  The lower panel of Figure 9 presents our result; the
uncertainties are given by Poisson statistics, and do not include the
systematic errors.  As expected, this distribution shows that the
effects of the correction are most important below $q\sim 0.3$.  For
concreteness, we refer to this as the corrected mass ratio
distribution.

Halbwachs et al.\ (2003) suggested that the mass ratio distribution is
different for the short and the long period spectroscopic
binaries. Trying different periods, they found the best period to
separate their sample into short and the long periods is 50 days.  We
searched for such a difference in our data following Halbwachs et al.\
approach. Our analysis indicated that if such an effect does exist in
our data, it is most pronounced when the sample is divided by a period
limit of 100 days. The resulting subsamples include 38 and 24 long and
short period binaries, respectively. The resulting mass ratio
distributions are depicted in Figure~10.

The two subsamples are smaller than the united sample, and
consequently their numerical noise is much larger. Therefore, we
derived the two distributions with only six bins. Further, we
omitted the first bin in the plot because its uncertainty in the
long period subsample is too large to allow any significant
comparison. The comparison between the five plotted bins suggests that
the mass ratio distribution of the long period binaries rises towards
low mass ratio, while the distribution of the short period binaries
does not. However, the small sample size and hence large uncertainties
in the low-q bins does not allow us to establish this finding firmly, 
and we need more binaries to investigate the possible difference.

\section{Comparison with Previous Studies}

Both mass ratio distributions shown in Figure 9, the directly measured
one (top panel) and the one corrected for undetected binaries (lower
panel), are approximately flat between $q=1$ and $\sim 0.3$.  This
result appears to differ from the recent derivations of
the mass ratio distribution obtained by Halbwachs et al.\ (2003) and
Tokovinin (2000), and is slightly different from the results of G03.

G03 analyzed the whole sample of SBs found in the CL94 survey, and
suggested that the mass ratio distribution of the Galactic disk
binaries rises monotonically toward low $q$'s.  Their Figure 6 (right
panel) shows, however, that the rise could be confined to $q<0.3$, and
that the distribution is consistent with having an approximately
constant value between $q =0.3$ and 1. The Galactic disk sample of
G03, consisting of 73 binaries, is very similar to ours, as both were
chosen from the CL94 survey. Their sample is somewhat larger, as they
did not exclude binaries with distances larger than 90 pc. On the other
hand, we added the information about the mass ratio for an additional 32
binaries. Therefore, the flatness of the distribution of the CL94
Galactic disk sample down to $q\sim 0.3$ seems firm from our results.
We suggest that the insignificance rise of the distribution in the
mass ratio range of 0.3--0.5 indicated by the analysis of G03 is the
result of the inherently low resolution of the analysis. This is
caused by the lack of mass ratio information for most of the binaries,
which can not be recovered by the statistical analysis.

Halbwachs et al.'s (2003) mass ratio distribution shows a prominent
peak near $q=1$. This feature confirms the Tokovinin (2000) study, which
found a large excess of ``twin binaries'' in the range of $q$ between
$0.95$ and 1. Both studies found this excess to be confined to
binaries with periods shorter than 30--50 days.  Tokovinin's sample of
SB2s with periods between 2 and 30 days included only 5 systems in the
mass ratio range of 0.8--0.9 and 41 binaries in the range of
0.9--1, an enhancement of a factor of $\sim 8$.  
We do not find such a peak, neither in our Figure~9, 
which presents our best estimate for the
mass distribution of the whole sample studied here, nor in Figure~10,
which separates our sample for short and long period binaries.

Tokovinin (2000) studied a sample of published SB2s, detected
by different spectrographs and observers. The sample
includes the SB2s listed in the Eighth Catalog of Spectroscopic
binaries (Batten et al.\ 1989) and the SB2s included in his
compilation of orbits published since then.  All the binaries were
detected either by direct measurements of the secondary lines, or by
resolving the two spectra with the one-dimensional correlation
function.   

The ability to detect the secondary is sensitive to the light ratio of the
primary and the secondary, which in turn strongly depends on the mass
ratio. For example, many of the binaries compiled by Tokovinin (2000) were
detected by the CfA spectrograph which operates in the visible light,
centered at 5200 \AA. At this band the stellar luminosity, $L_{5200}$,
depends on mass as $L_{5200}\propto M^{\sim 7.4\pm 0.6}$
(G02). Therefore, at q=0.85, for example, the light ratio falls down
to $\sim 0.3$, a value that makes the detection of the secondary
velocity difficult, depending on the S/N and resolution of the 
spectra. The fact that Tokovinin's sample includes a few SB2s with $q
\sim 0.8$ does not exclude a strong selection effect against finding
such binaries. Those binaries could have been observed with
exceptionally high S/N ratio, so that their secondary could have been
detected. Tokovinin (2000) noted that his sample could have also included the 
\"Opik effect.  We suggest that the high
peak at $q\sim1$ in Tokovinin's study could be the result 
of a the combination of the two selection effects.
The absence of a similar peak in Tokovinin's sample for the long-period
sample could have arisen in the intrinsic difficulty of resolving 
the two sets of lines in the spectra of the long period binaries.

We now consider whether our sample
could have suffered from an opposite selection effect that prevented
us from detecting binaries with $q$ between 0.95 and 1. We could have
missed such systems if L02 did not identify binaries with mass ratios
close to unity in the original survey for binaries in the sample of
high-proper-motion stars. L02 observed all stars in that sample a few
times and carefully examined the one-dimensional correlation function
of the measured spectra. To escape detection by the L02 survey a binary
had to fulfill two conditions. First, the one-dimensional correlation
function must not resolve the two set of lines, which can happen only
for binaries with small radial velocity amplitude. Second, the radial
velocity of the single, blended peak of the correlation must not vary.
The blended peak represented the spectral lines of the
primary and the secondary together, and therefore was sensitive to the
radial velocities of both spectra. Thus, the center of this peak would
not vary only if the contribution of the variation of the primary was
canceled out by the secondary. This could happen if the light ratio and
the radial velocity of the primary and the secondary were the same,
possible only in a very narrow range of mass ratios, between, say,
$q=0.98$ and unity.

In fact, there is a way to detect some binaries in this range of mass
ratios, as Halbwach's et al. (2003) noted.  One could search for the 
variation of the width of the peak of the one-dimension correlation. 
Halbwachs et al.\ (2003) did not include such ``line width spectroscopic 
binaries'' in their analysis, because they considered that ``the derivation 
of their orbital elements are rather hazardous''. The result, 
they commented, is a mass ratio distribution which is biased against twin 
binaries.

L02, however,  have scrutinized carefully the one-dimensional
correlation functions for peaks with variable widths, in order to
detect binaries hidden by this effect. Once a system with a variable
width of the one-dimensional correlation peak was detected, L02 and G02
thoroughly analyzed such systems with TODCOR to try to resolve the two
velocities. Obviously, some binaries with mass ratio between, say,
$q\sim 0.98$ and 1 and a small radial velocity amplitude could have
escaped L02 scrutiny. However, this effect would be the strongest in
long period binaries, in contrast to Figure~10, which does not
show the twin binary feature for both subsamples. We therefore suggest
that the effect in our sample is small and could not explain the
discrepancy between our results and those of Tokovinin (2000).

The peak at $q \sim 1$ measured by Halbwach's et al. (2003) is
much smaller than that of Tokovinin's sample. Halbwach's et al.  
Figure 7a suggests that the mass ratio frequency at the peak is higher by a 
factor of $2.7 \pm 0.7$ than the averaged frequency in the range of
0.2--0.9. This peak is based in part on the correction for the line
width spectroscopic binaries not included in their analysis. Without
the correction, the peak is only 1.4 higher than the averaged
frequency. This peak is still not consistent with our results, but the
difference is not very significant. Apparently, a much larger
sample in needed to establish whether this peak is real or not.

\section{Discussion}

The main result of our analysis is the flatness of the mass ratio
distribution between $q=0.3$ and $1$.
Between $q=0.1$ and $0.3$ the distribution might be higher than its
value at $q>0.3$, consistent with the peak found by G03 at $q \sim
0.2$.  However, the height of the possible rise cannot be determined
with confidence, because the uncertainty of the correction at low mass
ratios is large and depends on the true distribution of periods in the
sample.  For example, if we had assumed a flat distribution in $P$
rather than one flat in $\log P$, the correction would have been
larger, and the peak at low q's more prominent.  A determination of
the mass ratio distribution below $q\sim 0.3$ will require a radial
velocity survey with sensitivity better than the CL94 one (L02,
G02). We find some evidence that this peak is confined to the binaries
with periods longer than 100 days. Below $q=0.1$ our results are very
uncertain. 

The flatness of the mass ratio distribution between $q=0.3$ and $1$
can already put some constraints on the binary formation models. For
example, Bate, Bonnell \& Bromm (2002) suggested that a high frequency
of close binaries can be produced through a combination of dynamical
interaction in unstable multiple systems and orbital decay of
initially wider binaries. Orbital decay may occur as a result of gas
accretion and/or the interaction of a binary with its circumbinary
disc. They found that such mechanisms result in a preference for close
binaries to have roughly equal-mass components, a tendency that is not
found by our results.  On the other hand, dynamical disintegration of
small clusters of stars could result in very different mass ratio
distributions, depending on the exact parameters of the dynamical
model used (Durisen, Sterzik, \& Pickett 2000). Durisen, Sterzik, \&
Pickett (2001) found one model which produced for G type primaries
almost flat distribution.

To place our findings in a wider context, we consider the distribution
of the masses of the secondaries in our sample.  For a sample as
homogeneous as ours, the distribution of mass ratios reflects that of
secondary masses. Using the median primary mass of 0.72 \Mo, our
result indicates a secondary mass spectrum that is approximately flat
down to 0.2 \Mo, and possibly higher between 0.07--0.2 \Mo.  This
differs from the monotonic rise toward low masses found in the single
star Initial Mass Functions (IMFs) of young open clusters (e.g.,
Barrado y Navascues 2002; Tej et al.\  2002; Moreaux et al.\ 2003) and
the much younger star forming regions (Luhman et al.\ 2000).  It also
differs from the log-normal IMF derived by Chabrier (2003) for the
galactic field stars.  Our results show that the secondaries in
binaries with periods $P< 3000$ days are not captured from a
single-star IMF, and that their formation mechanisms must differ from
that for single stars.

It is interesting to compare the distribution of the masses of stellar
secondaries with that of the extra-solar planets.  In order to compare
the two distributions we need to scale their respective parent samples
to ones of same size.  The binaries studied in this work are found in
a set of 488 disk stars closer than 90 pc within the larger CL94
sample. The size of the parent sample of the extra-solar planets is
not known.  We assume, somewhat arbitrarily, that the planets came
from a parent sample of 2000 stars.  We scaled both to a common sample
size of 1000.  The results are shown in Figure~11; the top panel plots
the secondary mass distribution of our sample, and the lower panel the
mass distribution of the extra-solar planets, both on a logarithmic
scale in mass.  The scale of the ordinate in both panels is
$dN_{1000}/d\log M_2$, where $N_{1000}$ is the cumulative number of
secondaries up to $M_2$ for a parent sample of 1000 single (and
multiple) stars.  We plotted the secondary distribution down to 0.07
\Mo, which corresponds to $q=0.1$; our measurements are very
insensitive below this value.  (While Figure~9 presents number of
binaries per mass ratio bin, Figure~11 gives the number per log $M_2$
increment.)

To calculate the mass distribution of the extra-solar planets, we used
the compilation of the 101 planets as of April,
2003\footnote{http://exoplanets.org/almanacframe.html}, and applied
the MAXLIMA algorithm of Zucker \& Mazeh (2001), which derives the
distribution of masses from their minimum values, the values actually
measured. Following Zucker \& Mazeh, we assumed that the population of
planets is complete for those with velocity semi-amplitude greater
than 40 m s$^{-1}$. We therefore used in our analysis only the 70
planets around G and K primaries, with $K_1$ larger than this
threshold, and corrected the distribution accordingly. The dashed line
shows the uncorrected distribution. The uncertainties of the corrected
distribution represent the random errors, derived with Monte-Carlo
simulations. They do not represent the systematic errors that might be
the consequence of the assumed period distribution or even the basic
assumption of the independence of the mass and period distributions
(e.g., Zucker \& Mazeh 2002).  The scale of the two panels is
different because we find that the frequency of stellar secondaries in
the 0.07--0.7 \Mo\ range, with $P< 3000$ d, is substantially larger
than the presently detected frequency of extra-solar planets in the
1--10 Jupiter mass ($10^{-3}$--$10^{-2}$ \Mo) range. The former is
about 15\%, whereas the latter is about 3\%, assuming a parent sample
of 2000 stars.

The two panels suggest the presence of two distinct populations.
Plotted on a logarithmic basis in mass, the stellar secondary
population decreases toward the regime of the brown dwarfs.  The
planetary distribution is consistent with a cutoff above $\sim 15$
Jupiter masses and a flat distribution below this value. The gap
separating the two populations is the ``brown-dwarf desert" which
suggests a natural distinction between a planet and a low-mass stellar
secondary based on mass (e.g., Jorissen, Mayor \& Udry 2001; Zucker \&
Mazeh 2001). The decline of the stellar population towards the
brown-dwarf region is consistent with the general trends found by
Luhman et al.\ (2000) and Moraux et al.\ (2003).

\section{Closing Remarks}

The present analysis of spectroscopic binaries, which relies on
combining systematic, precision spectroscopy in visible and IR light,
shows that for solar-type main-sequence stars it is already possible to
measure the mass ratio distribution in close binaries down to $q\sim$
0.2--0.3, thereby providing empirical input to theories of binary
formation with minimal statistical inference.  The prospects are
excellent for improving the sensitivity limit to lower mass ratios by
analyzing several IR spectroscopic orders simultaneously (Bender et
al.\ 2003).

Purely spectroscopic observations of SB2s, however, leave the orbital
inclination, and hence component masses, undetermined.  The
inclination can be measured by combining the spectroscopic results
with interferometric observations that resolve the binary.  Boden,
Creech-Eakman, \& Queloz (2000) have illustrated this approach to
derive precise component masses using the Palomar Testbed
Interferometer, and Halbwachs et al.\ (2003) used {\it HIPPARCOS}
measurements for the same purpose in their study of nearby SBs.  The
development of ground-based IR interferometers, and the advent of the
{\it Space Interferometry Mission}, offer the promise of significant
progress in this area in the coming decade.

\begin{acknowledgements}

We thank the referee for a careful review and very helpful comments
that helped us to improve the manuscript. We are grateful to Dorit
Goldberg and Dara Norman for their assistance in the early stages of
this project.  This research was supported in part by NSF Grant
02-05427 (MS), by NASA through the American Astronomical Society's
Small Research Grant Program (LP), and by the Israeli Science
Foundation grant no.\ 40/00 (TM).  SZ is grateful for partial support
from the Jacob and Riva Damm Foundation.  Data presented herein were
obtained at the W.M. Keck Observatory, which is operated as a
scientific partnership between the California Institute of Technology,
the University of California, and NASA.  The Observatory was made
possible by the generous financial support of the W.M.\ Keck
Foundation.  The authors extend special thanks to those of Hawaiian
ancestry on whose sacred mountain we are privileged to be guests.

\end{acknowledgements}

\clearpage

\begin{figure}
    \plotone{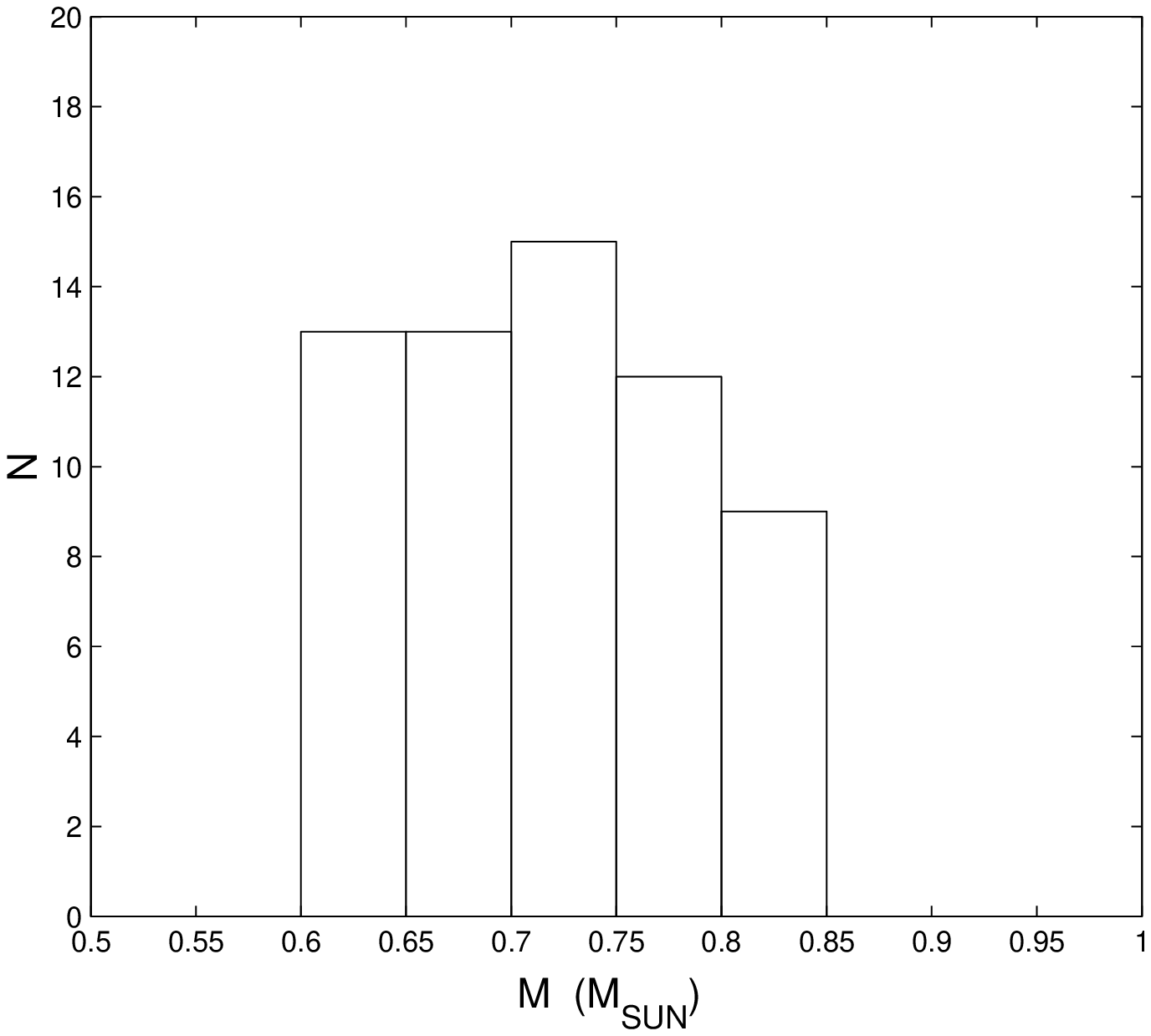} 
\medskip 
\caption{The distribution of
    primary masses, $M_{1,est}$, as estimated by CL94, in the entire
    sample of 62 binaries analyzed in this study --- 11 SB2s
    identified in visible light by GO2, and 51 SB1s observed by high
    resolution IR spectroscopy.}
\end{figure}

\begin{figure}
    \plotone{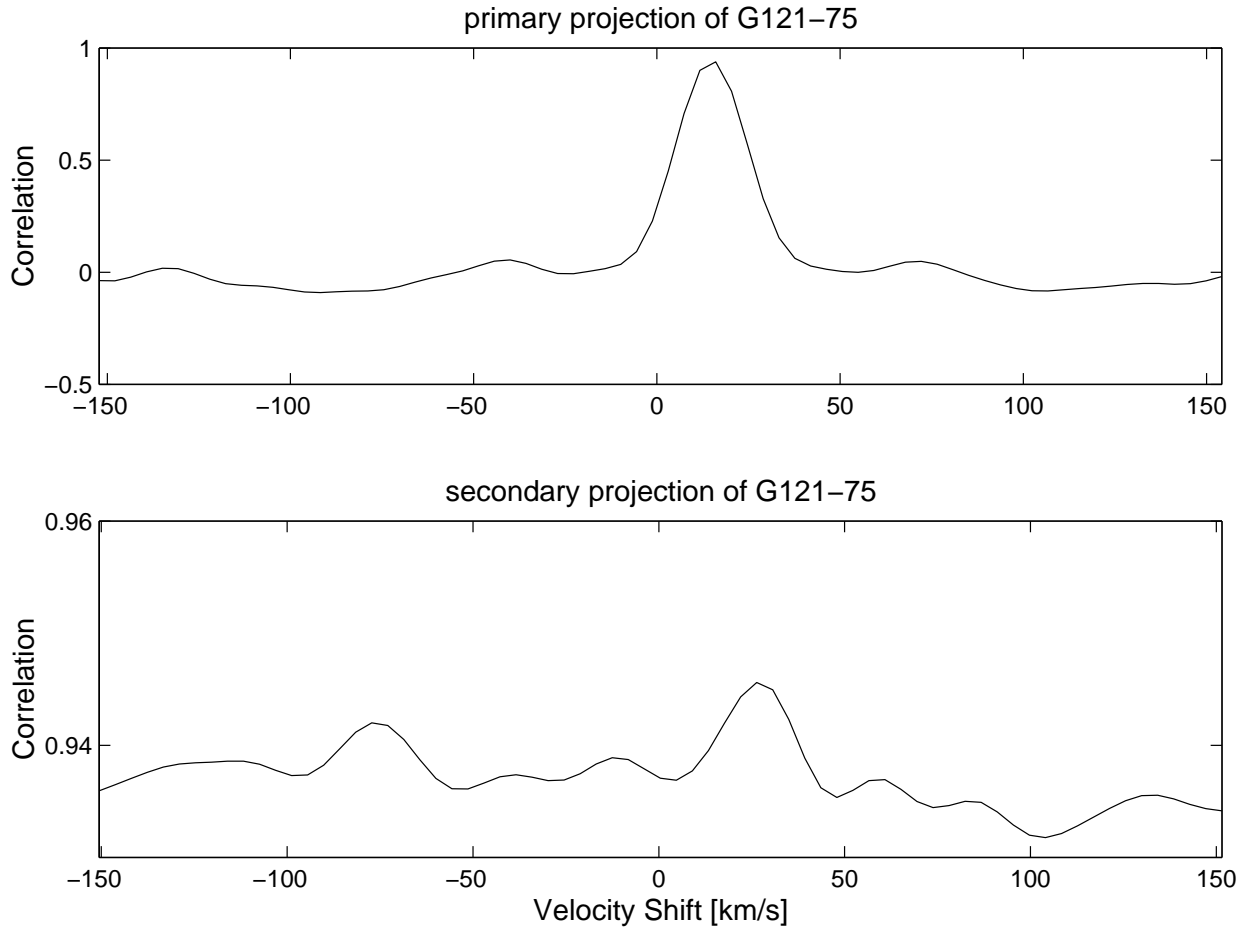}
    \medskip
    \caption{TODCOR output for the spectrum of SB2 G121-75, obtained
on 2 May, 2001 (UT), exhibiting two aspects of the two-dimensional
correlation function. The top panel shows the correlation as a
function of the primary template shift, when the velocity of the
secondary template is held fixed at the value that yielded maximum
correlation. The lower panel shows the correlation as a function of
the secondary template shift, when the velocity of the primary
template is held fixed at the value that yielded maximum
correlation. Note the different scales of the two panels.}
\end{figure}

\begin{figure}
    \plotone{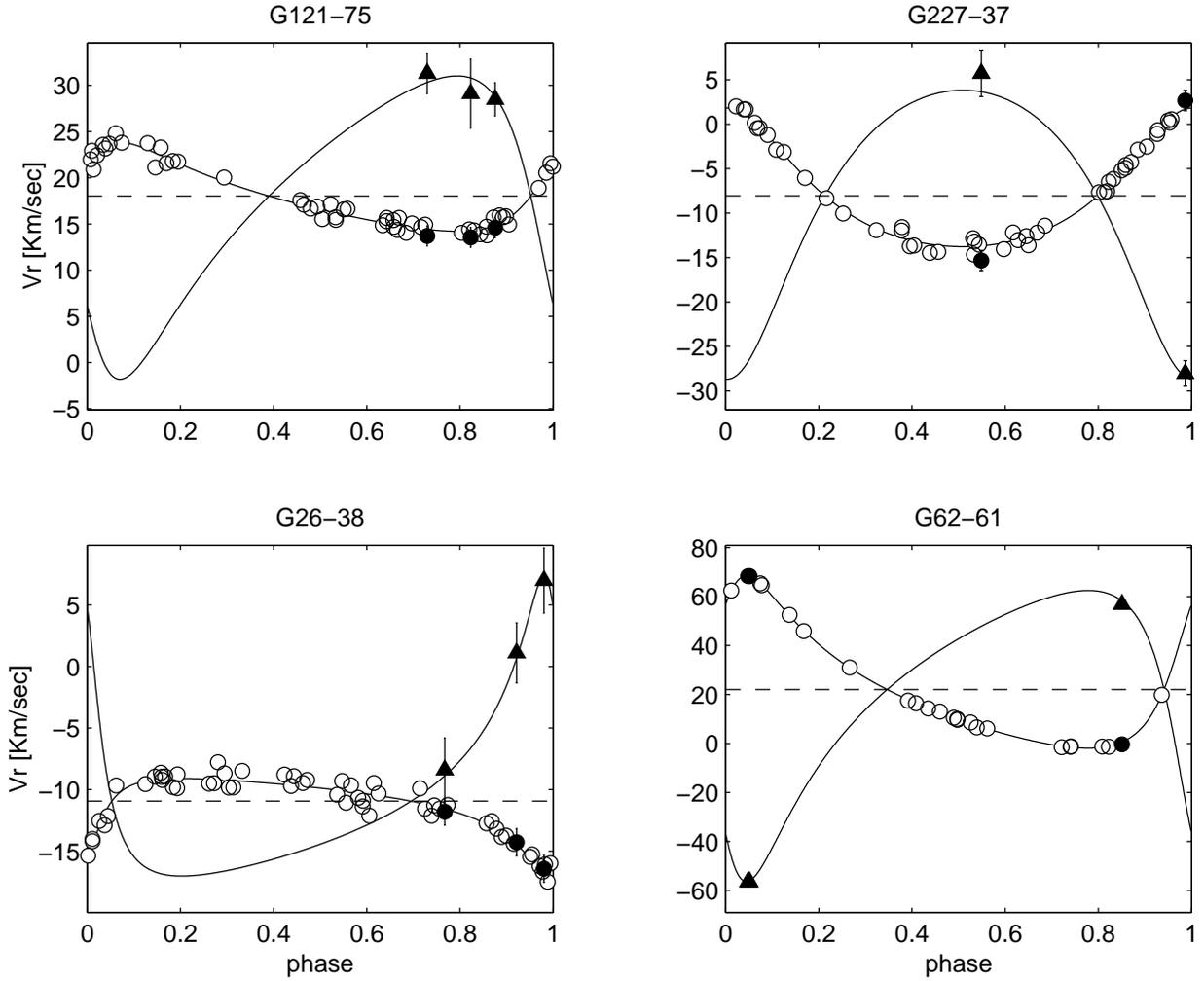}
    \medskip
    \caption{Double-lined orbital solution for four SB2s.
The open circles present previously measured velocities of the
      primary (L02).  The filled circles and triangles mark the
      primary and secondary velocities measured by this work. The
      solid lines present the calculated orbital solutions. }
\end{figure}

\begin{figure}
    \plotone{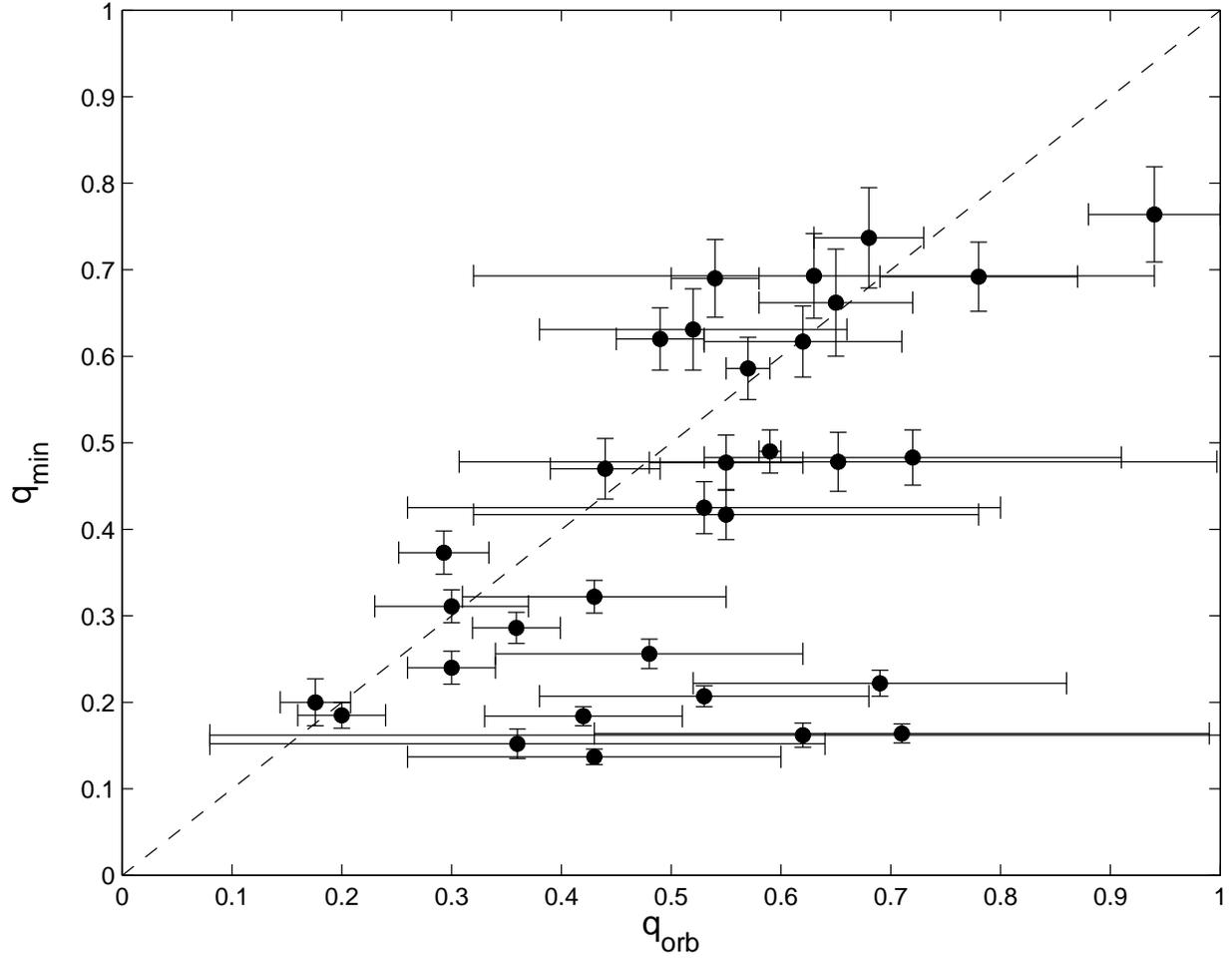}
    \medskip
    \caption{Minimum mass ratio as a function of the  measured mass ratio 
for the 32 SB2s in which IR spectroscopy detected the secondary. The
dashed line equation is $q_{min}=q_{orb}$.}
\end{figure}

\begin{figure}
    \plotone{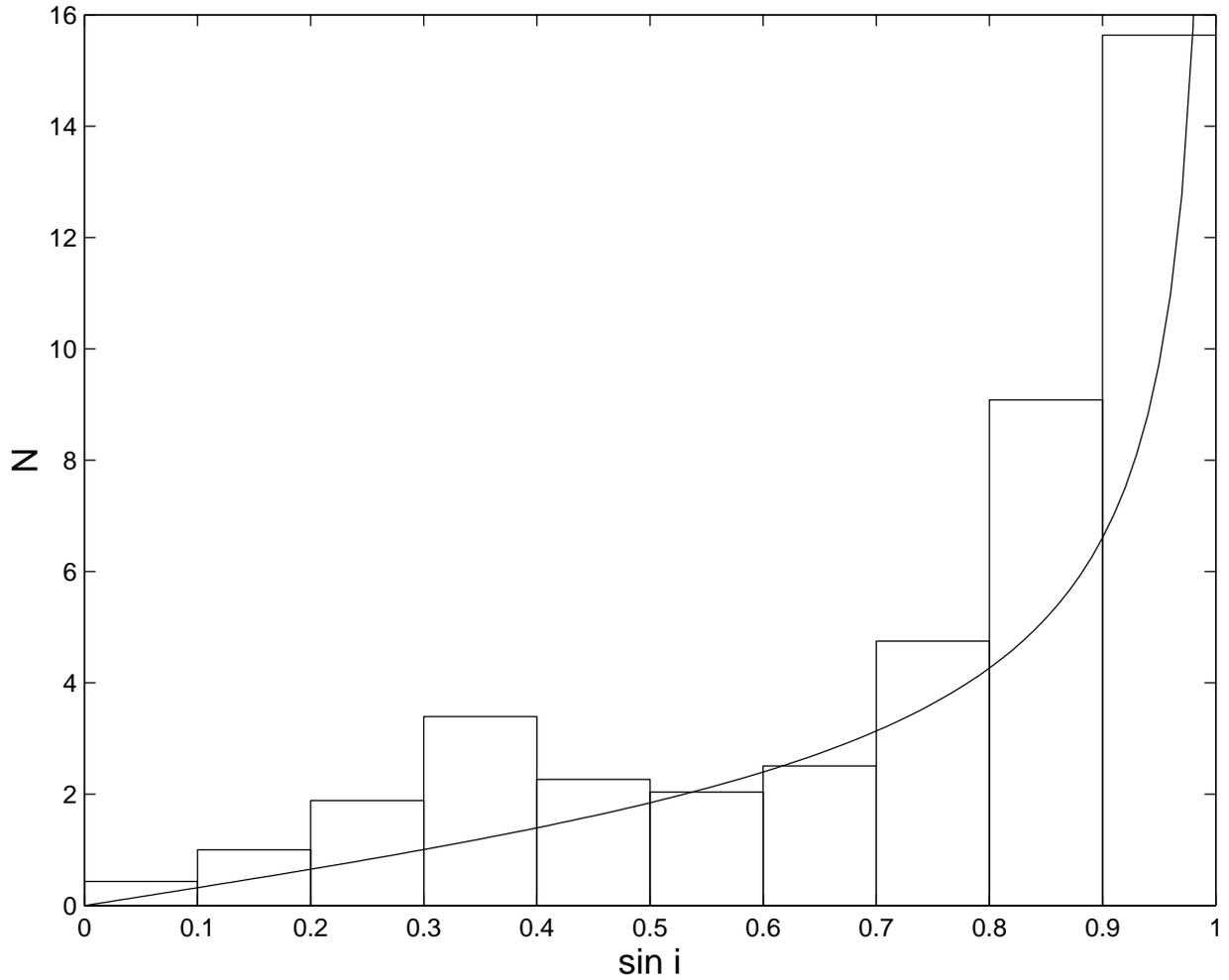}
    \medskip
    \caption{Histogram of derived 
$\sin~i$, as calculated from the mass function, $q_{orb}$,
and $M_{1,est}$ for the 32 SB2s detected by IR spectroscopy
 and the 11 SB2s detected by G02. The solid
line shows the expected distribution for randomly oriented orbits.}
\end{figure}

\begin{figure}
    \plotone{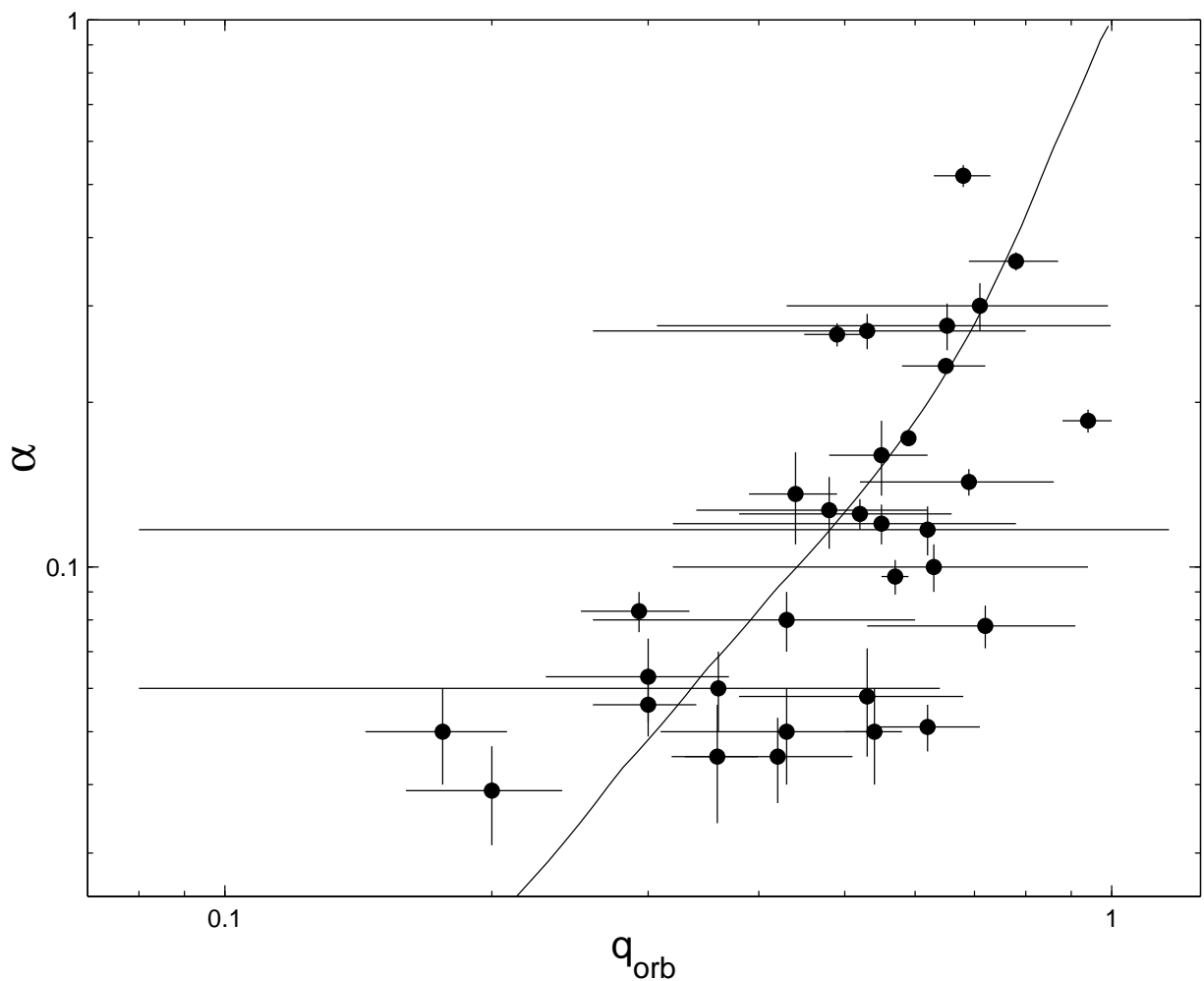}
    \medskip
    \caption{The secondary/primary flux ratio at $1.55\mu$m, $\alpha$, vs.\
$q_{orb}$ for the SB2s detected by IR spectroscopy.  The solid line shows
the theoretical $H$-band
flux ratio vs.\  mass ratio according to the calculations
of Baraffe et  al. (1998) for $M_1 = 0.7$ \msun, [Fe/H] = 0, and Age 5 Byr.}
\end{figure}

\begin{figure}
    \plotone{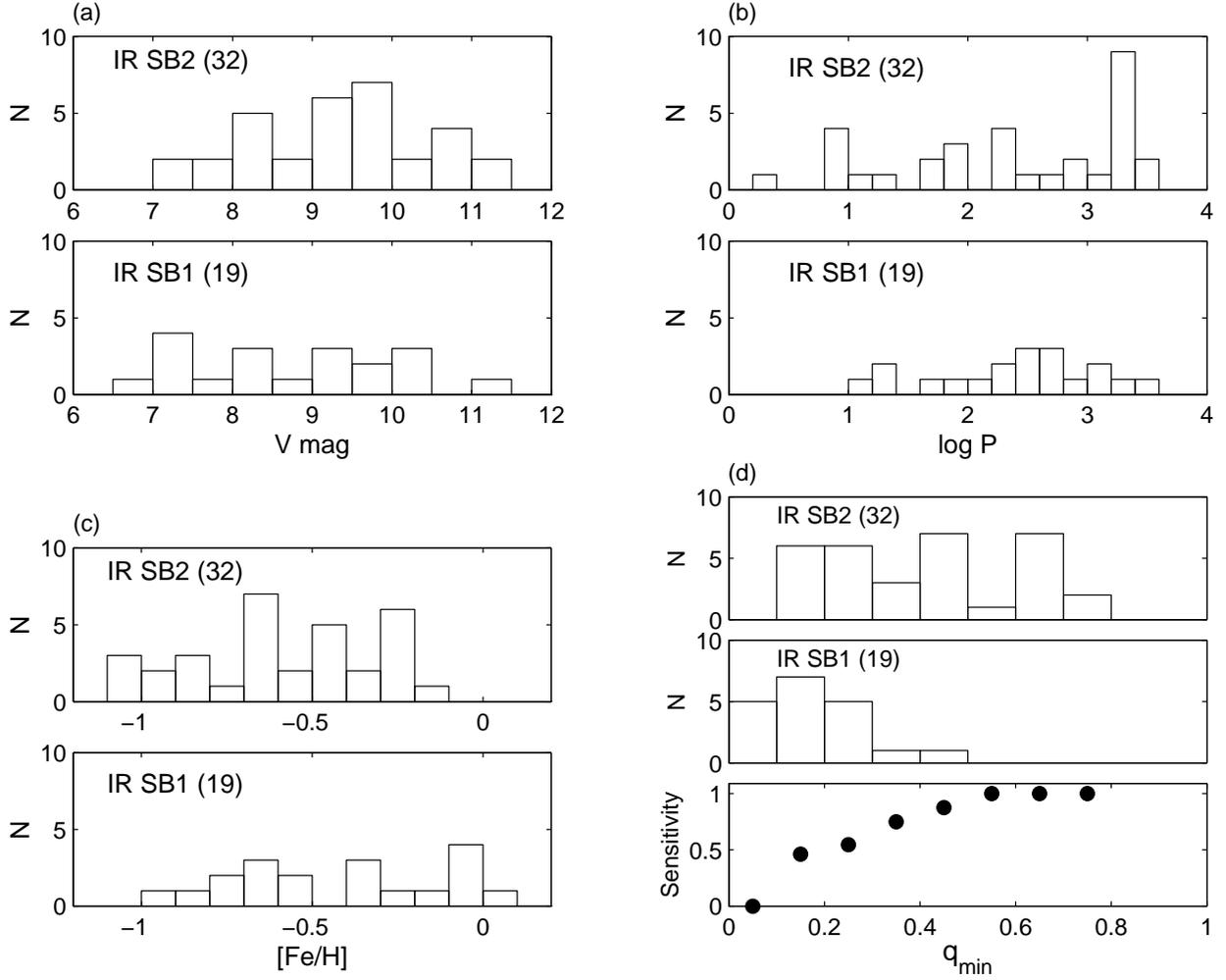}
    \medskip
    \caption{Comparison of the distributions of the $V$-band magnitude
(upper left), period (upper right), metallicity (lower left), and minimum 
mass ratio (lower right) for the 32 SB2s in which the secondaries were 
detected by IR spectroscopy, and for  the 19 binaries in which  the 
secondaries remain undetected.  In the distribution with respect to the 
minimum mass ratio, the bottom panel shows the sensitivity for detection given
as (number detected/total number) for each bin of
width $\Delta q_{min} = 0.1$.}
\end{figure}

\begin{figure}
    \plotone{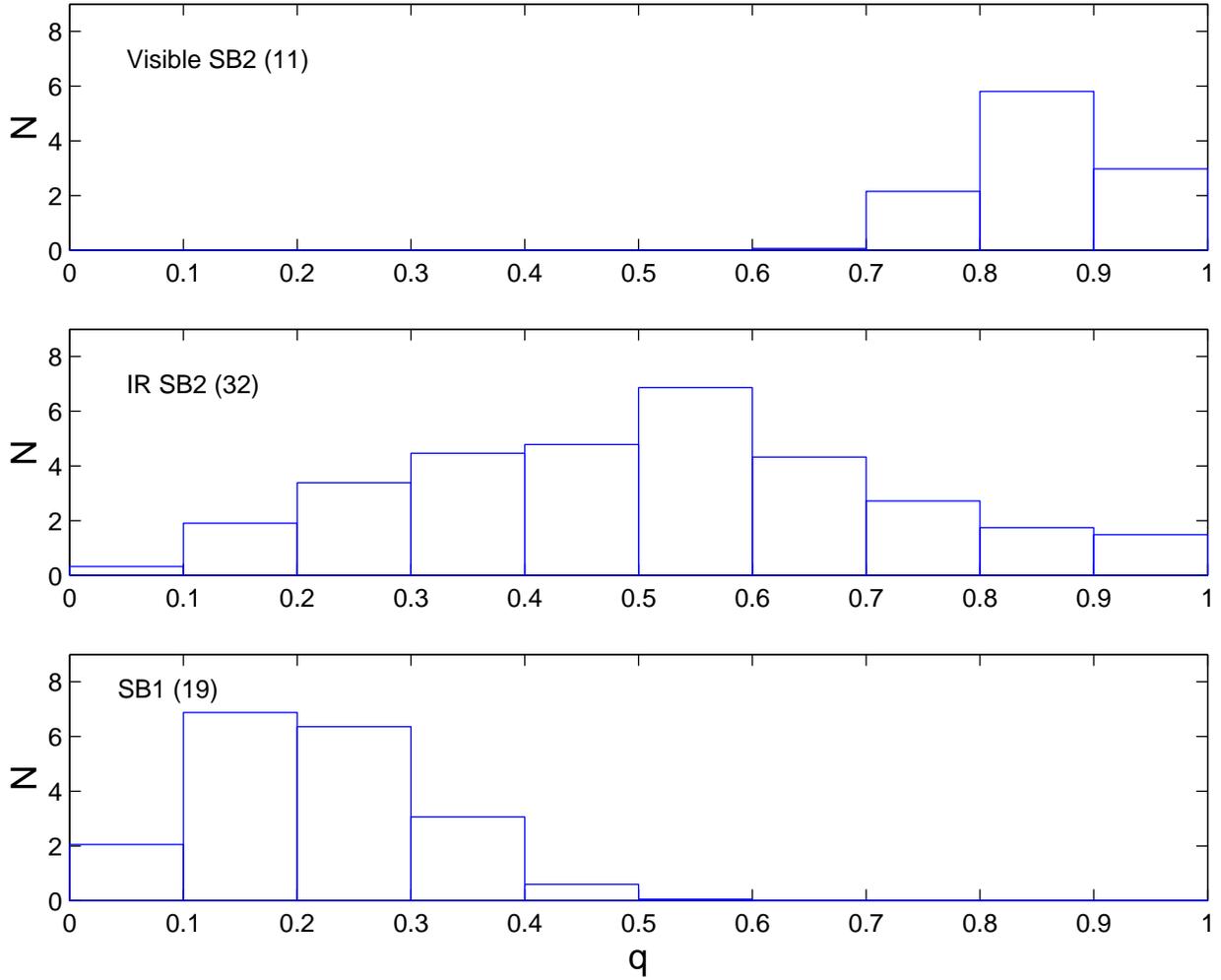}
    \medskip
    \caption{The distribution of mass ratios in our sample
in bins of width $\Delta q=0.1$.  Top panel --- measured in
visible light (G02), middle panel --- measured in the IR (this work),
and bottom panel --- the distribution for the 19 SB1s remaining in 
our sample, estimated with Mazeh and Goldberg (1992) algorithm
(see text, and compare with middle panel of Figure 7d).}
\end{figure}

\begin{figure}
    \plotone{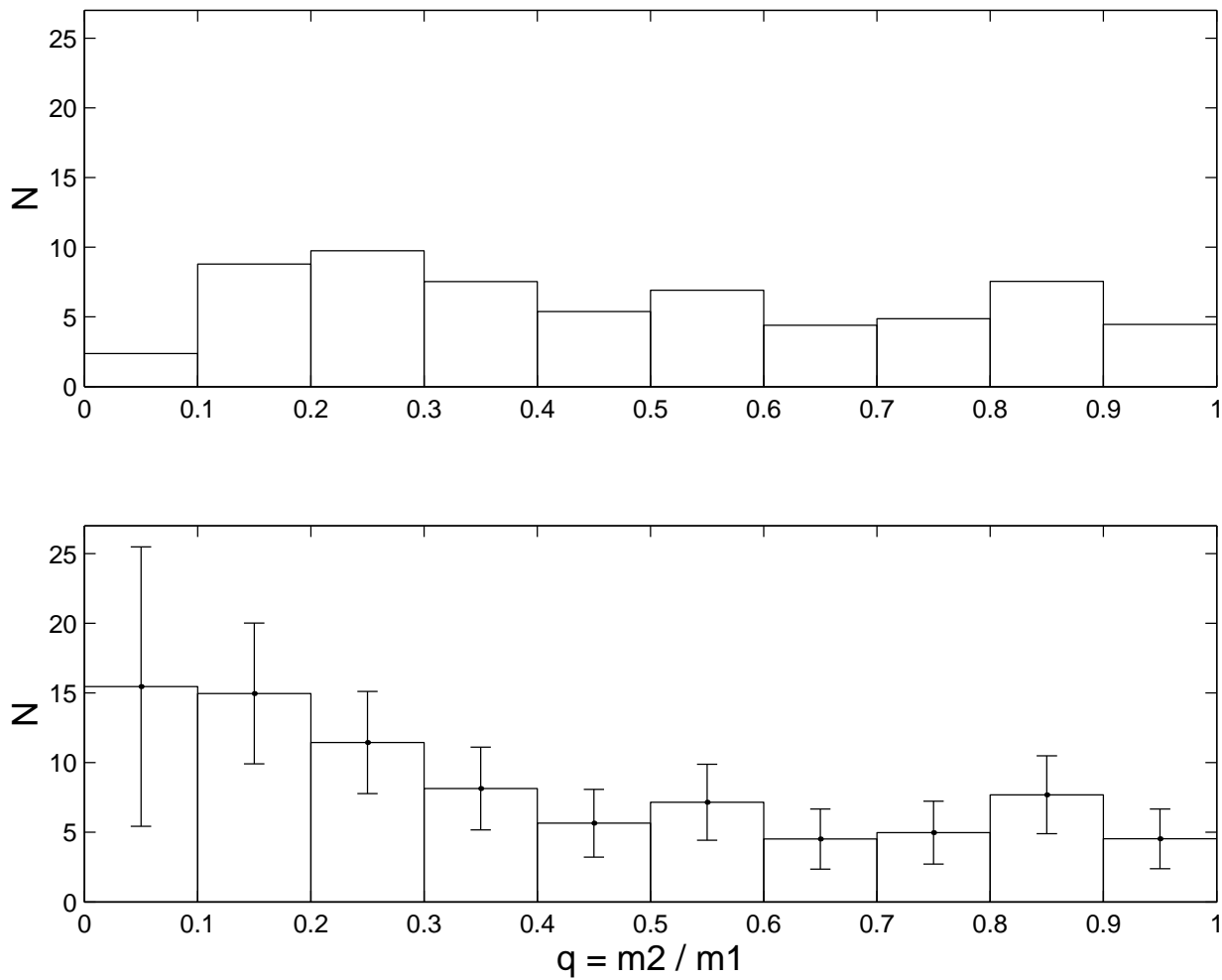}
    \medskip
    \caption{The derived mass ratio distribution plotted as a function
of $q$. {\it Top panel:} The sum
of the mass ratio distributions in the three panels of Figure 8.  This
is our estimate of the {\it directly} measured mass ratio distribution
of our sample; it should be reliable above $q \sim 0.3$. {\it Bottom
panel:} The distribution of the top panel, corrected for binaries
undetected in the original L02 sample (see text).  The corrections are
important below $q\sim 0.3$.}
\end{figure}

\begin{figure}
  \plotone{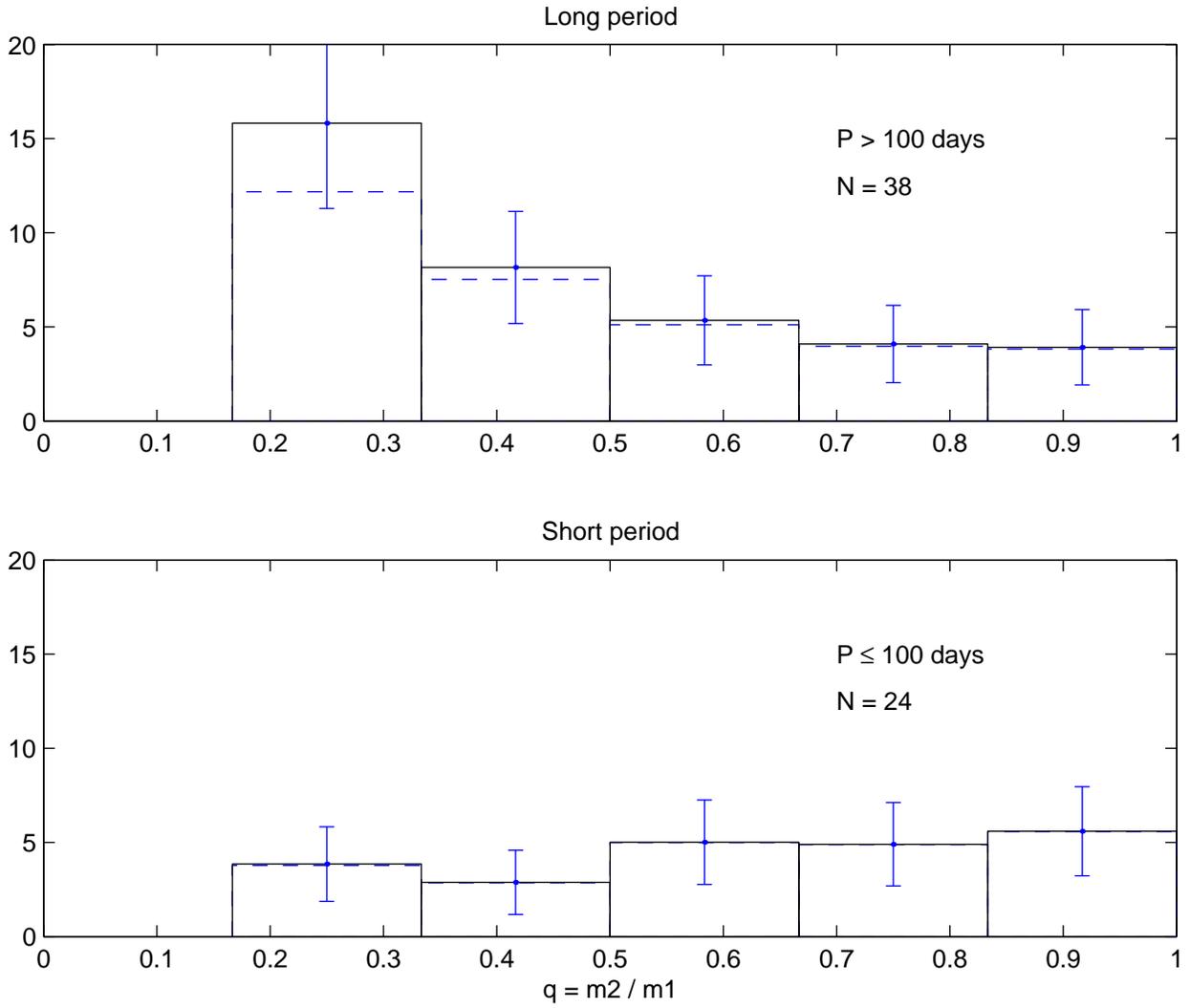} \medskip
    \caption{ The mass ratio distribution of the short and the long
    period binaries in our sample.}
\end{figure}

\begin{figure}
  \plotone{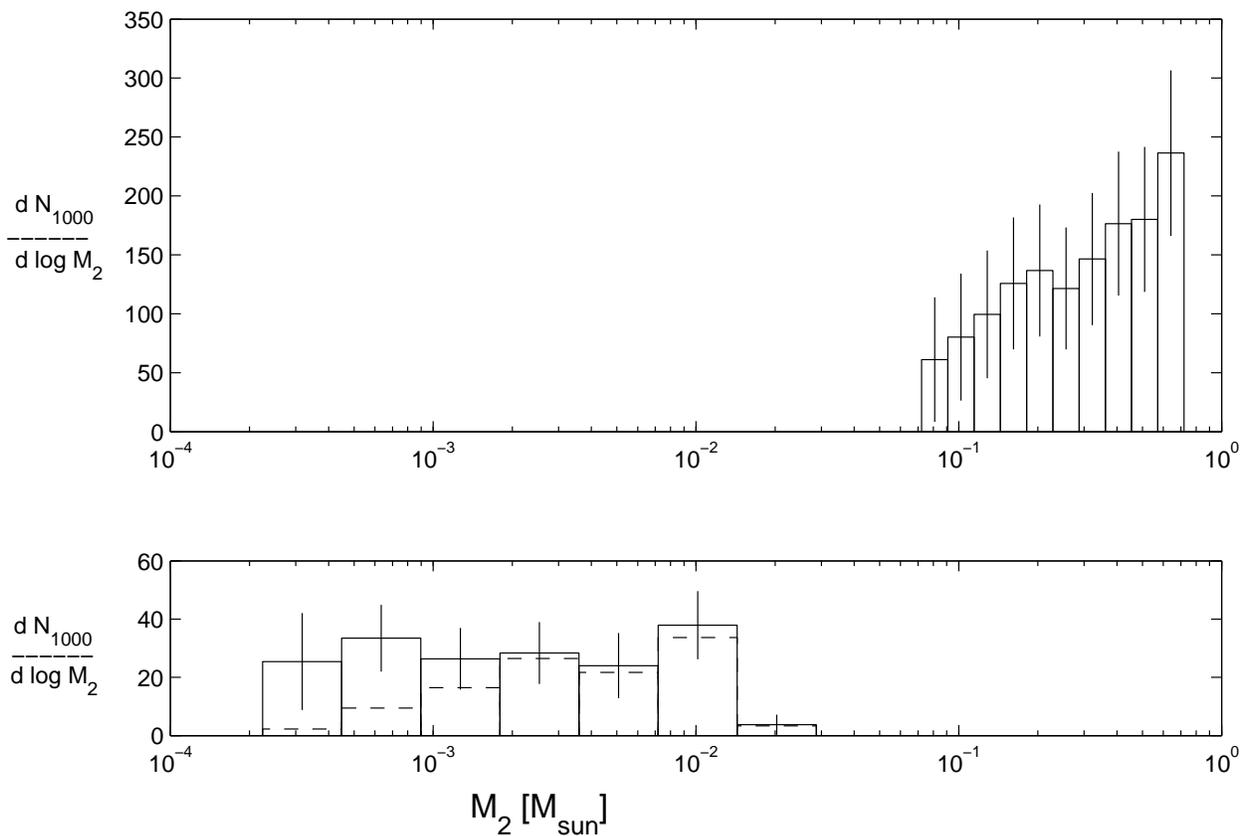} \medskip
    \caption{ The mass distribution of the stellar
      secondaries and the extra-solar planets. Both panels give the
      number of companions for a sample of 1000 single (some of which
      are multiple) stars per logarithmic unit of secondary mass (see
      text). The top panel is based on the present work, which is
      insensitive below 0.07 \Mo. The bottom panel is based on the
      present compilation of the published planets (dashed line)
      corrected for the undetected planets (solid line). The scales of
      the two panels are different.}
\end{figure}

\clearpage \pagestyle{empty}

\begin{deluxetable}{lllrrrrlc}
\tablewidth{0pt}
\tablecaption{Single-Lined Binaries in Sample\label{tbl-1}}
\tablehead{
\colhead{System} &\colhead{RA(2000)}  & \colhead{Dec(2000)}&
\colhead{V}  &\colhead{[Fe/H]} &\colhead{d} &\colhead{P}
&\colhead{M$_{1,est}$}
&\colhead{$q_{min}$}\\
\colhead{$~$}&\colhead{$~$}&\colhead{$~$}&\colhead{(mag)}&\colhead{$~$}&
\colhead{(pc)}&\colhead{(days)}&\colhead{(\msun)}  }
\startdata
G130-32&00:00:03.4&+34:11:22& 8.50&-0.58   & 41  & 951.\pzz& 0.71  &$0.16\pm0.01$\\
G32-49 &00:47:36.5&+14:38:22&10.94&-0.49   & 77  &  20.71 &  0.65 &$0.29\pm0.02$\\
G173-2 &01:30:55.3&+52:44:43&10.05&-0.07   & 64  & 339.79 &  0.79 &$0.16\pm0.01$\\
G72-58 &02:08:23.8&+28:18:38& 9.99&-0.55   & 61  & 208.0\pz &  0.67 &$0.23\pm0.02$\\
G72-59 &02:08:23.8&+28:18:17&10.55&-0.59   & 61  &  87.75 &  0.63 &$0.48\pm0.03$\\
G173-59&02:33:53.7&+49:30:22& 7.75&-0.50   & 33  &2323.\pzz&  0.75 &$0.62\pm0.04$\\
G78-1  &02:41:45.7&+47:21:01& 9.16&-0.86   & 67  & 183.41 &  0.73 &$0.14\pm0.01$\\
G77-56 &03:29:18.6&+01:58:31&10.45&-0.43   & 88  & 179.00 &  0.73 &$0.26\pm0.02$\\
G6-20  &03:37:11.0&+25:59:27& 7.26&-0.25   & 37  &   1.93 &  0.80 &$0.69\pm0.04$\\
G248-27&04:56:36.2&+72:57:05& 9.89&-0.61   & 52  &   7.53 &  0.66 &$0.18\pm0.01$\\
G84-39 &05:12:45.2&+04:19:16&10.58&-0.77   & 72  &   8.66 &  0.63 &$0.48\pm0.03$\\
G249-37&06:06:24.6&+63:50:06& 8.38&-0.27   & 26  &  48.96 &  0.73 &$0.32\pm0.02$\\
G87-20 &07:03:04.8&+38:08:32& 9.45&-0.96   & 57  &  85.18 &  0.66 &$0.76\pm0.06$\\
G108-53&07:05:04.1&+01:23:50& 8.45&-0.68   & 41  & 612.0\pz &  0.71 &$0.69\pm0.05$\\
G88-5  &07:06:08.1&+18:38:10&10.21&-0.57   & 76  & 903.9\pz &  0.69 &$0.21\pm0.02$\\
G88-11 &07:10:37.3&+20:26:27& 9.09&-0.33   & 56  & 352.68 &  0.79 &$0.20\pm0.01$\\
G112-54&07:54:34.1&$-$01:24:44& 7.43&-0.95   & 17  & 451.5\pz &  0.62&$0.27\pm0.02$ \\
G40-5  &08:04:34.6&+15:21:51& 8.48&-0.07   & 32  &  75.88 &  0.79 &$0.19\pm0.02$\\
G40-8  &08:08:54.3&+24:37:26& 9.65&-1.02   & 54  & 733.5\pz &  0.64 &$0.22\pm0.02$\\
G9-42  &09:00:47.4&+21:27:12& 8.80&-0.16   & 38  & 120.30 &  0.77 &$0.05\pm0.01$\\
G41-34 &09:22:46.9&+11:16:18& 9.65&-0.20   & 60  & 193.82 &  0.78 &$0.15\pm0.02$\\
G195-30&09:30:18.7&+52:09:11& 9.46&-0.69   & 69  &  16.91 &  0.72 &$0.12\pm0.01$\\
G58-23 &10:49:52.6&+20:29:29& 9.96&-1.09   & 71  &1779.\pzz&  0.66 &$0.66\pm0.06$\\
G44-45 &10:53:23.7&+09:44:21&10.16&-0.33   & 64  &1837.\pzz&  0.71 &$0.40\pm0.03$\\
G147-36&11:17:14.5&+29:34:13& 9.29&-0.61   & 62  &   6.57 &  0.73 &$0.59\pm0.04$\\
G147-58&11:30:22.3&+35:50:30& 9.89&-0.64   & 74  &   7.15 &  0.71 &$0.69\pm0.05$\\
G59-5  &12:13:27.7&+23:15:56& 9.40&-0.80   & 64  &2483.\pzz&  0.70 &$0.43\pm0.03$\\
G121-75&12:19:00.7&+28:02:52& 8.96&-0.24   & 46  &2161.\pzz&  0.78 &$0.37\pm0.03$\\
G60-47 &12:55:15.9&+07:49:57& 9.93&-0.48   & 60  &1042.5\pz &  0.68 &$0.42\pm0.03$\\
G14-54 &13:28:18.7&$-$00:50:24& 7.43&-0.22   & 30  & 207.32 &  0.84 &$0.19\pm0.02$\\
G62-44 &13:31:39.9&$-$02:19:03& 7.34&-0.69   & 17  &1189.2\pz &  0.65 &$0.16\pm0.01$\\
G165-22&13:38:01.9&+39:10:41& 7.78&-0.23   & 32  &  11.58 &  0.82 &$0.09\pm0.01$\\
G62-61 &13:40:26.9&+02:09:05& 8.21&-0.42   & 48  &  14.49 &  0.83 &$0.49\pm0.03$\\
G255-45&13:56:15.1&+74:42:44& 9.72&-0.62   & 83  &1620.7\pz &  0.73 &$0.20\pm0.03$\\
G239-38&15:00:26.9&+71:45:55& 6.65&-0.68   & 18  & 468.1\pz &  0.71 &$0.11\pm0.01$\\
G15-6  &15:04:59.5&+04:05:17& 9.87&-0.76   & 63  & 265.65 &  0.67 &$0.10\pm0.01$\\
G66-65 &15:07:46.5&+08:52:47& 8.27&-0.89   & 38  & 201.71 &  0.70 &$0.06\pm0.01$\\
G202-25&15:59:56.3&+45:44:10&11.04&-0.38   & 88  &2727.\pzz&  0.69 &$0.33\pm0.02$\\
G17-22 &16:32:51.6&+03:14:45& 8.84&-0.88   & 25  & 226.21 &  0.59 &$1.36\pm0.13$\\
G227-37&18:35:09.3&+63:41;47& 8.07&-0.33   & 45  &1730.6\pz &  0.84 &$0.62\pm0.04$\\
G21-20 &18:37:58.8&$-$06:48:19& 8.34&-0.10   & 31  &  21.00 &  0.79 &$0.13\pm0.01$\\
G22-7  &18:55:52.9&$-$05:44:42& 7.46&-0.03   & 25  &  52.78 &  0.84 &$0.21\pm0.01$\\
G262-14&20:28:27.9&+62:00:52&11.46&-0.93   & 90  &  81.20 &  0.61 &$0.74\pm0.06$\\
G230-45&20:40:16.8&+54:13:14&11.43&-0.87   & 88  & 398.3\pz &  0.61 &$0.47\pm0.04$\\
G262-32&20:59:01.4&+65:02:43&10.73&-0.97   & 61  &  52.92 &  0.61 &$0.16\pm0.01$\\
G26-38 &21:58:45.1&+00:48:35&10.29&-0.29   & 79  &2660.\pzz&  0.76 &$0.24\pm0.02$\\
G18-55 &22:32:48.5&+10:24:17& 9.35&-0.68   & 39  &2916.\pzz&  0.64 &$0.63\pm0.05$\\
G67-38 &22:59:19.4&+12:11:32& 8.38&-0.67   & 44  &1623.\pzz&  0.73 &$0.48\pm0.03$\\
G156-75&23:01:51.5&$-$03:50:55& 7.43&+0.09   & 17  & 468.1\pz &  0.81 &$0.09\pm0.01$\\
G157-21&23:09:04.0&$-$02:33:01& 9.27&-0.76   & 54  &1134.\pzz&  0.68 &$0.23\pm0.02$\\
G68-31 &23:36:06.0&+18:26:33& 7.66&-0.35   & 28  &1810.\pzz&  0.76 &$0.31\pm0.02$\\
G130-10&23:46:09.4&+35:14:37& 9.13&-0.27   & 47  &1754.6\pz &  0.76 &$0.21\pm0.01$\\
\enddata
\end{deluxetable}

\clearpage
\pagestyle{empty}
\begin{deluxetable}{lllrrrrlc}
\tablewidth{0pt}

\tablecaption{SB2s Identified in Visible Light\label{tbl-2}}
\tablehead{
\colhead{System} &\colhead{RA(2000)}  & \colhead{Dec(2000)}&
\colhead{V}  &\colhead{[Fe/H]} &\colhead{d} &\colhead{P}   &\colhead{M$_{1,est}$}   &\colhead{$q$} \\
\colhead{$~$}&\colhead{$~$}&\colhead{$~$}&\colhead{(mag)}&\colhead{$~$}&
\colhead{(pc)}&\colhead{(days)}&\colhead{(\msun)}  }

\startdata

G69-1  &00:32:33.9&+28:11:51 &8.72&-0.62 &72   &2127.82&0.72 &$0.93\pm0.02$\\
G69-4S &00:36:02.3&+29:59:35 &8.46&-0.80 &40   &35.98  &0.67 &$0.82\pm0.01$\\
G34-39 &01:37:25.0&+25:10:04 &6.97&-0.51 &14   &25.21  &0.64 &$0.78\pm0.01$\\
G133-57&01:57:34.7&+42:04:36 &9.01&-0.22 &64   &694.87 &0.80 &$0.76\pm0.04$\\
G37-10 &02:53:01.4&+35:38:13 &8.38&-0.76 &37   &28.95  &0.63 &$0.89\pm0.01$\\
G6-26A &03:39:33.5&+18:23:05 &8.28&-0.89 &31   & 8.65  &0.61 &$0.96\pm0.01$\\
G59-32 &12:40:07.0&+20:48:32 &8.98&-0.23 &57   &31.02  &0.75 &$0.91\pm0.01$\\
G16-9  &15:45:52.4&+05:02:26 &9.15&-0.88 &40   & 9.94  &0.60 &$0.82\pm0.02$\\
G182-7 &17:24:42.4&+38:02:10 &8.60&-0.69 &36   &448.61 &0.64 &$0.88\pm0.02$\\
G209-35&20:32:51.6&+41:53:54 &7.08&-0.55 &21   &57.32  &0.65 &$0.84\pm0.03$\\
G210-46&20:59:55.2&+40:15:31 &6.57&-0.40 &35   &112.55 &0.84 &$0.85\pm0.02$\\
\enddata
\end{deluxetable}

\clearpage

  \pagestyle{empty}
  \begin{deluxetable}{lcrcrc}
  \tablewidth{0pt}
  \tablecaption{Observations and Measured Velocities\label{tbl-3}}
\tablehead{\colhead{System}& \colhead{$T_{obs}$}  &   \colhead{$v_{1}$}&\colhead{$\sigma$}  &
\colhead{$v_{2}$}&\colhead{$\sigma$} \\
\colhead{$ $}&\colhead{(MJD-50000)}&\colhead{(\kms) }&\colhead{(\kms)}&
\colhead{(\kms)}  &\colhead{(\kms)} } 
\startdata

G130-32  & 1916.7 &  -29.6 &   0.1  &    -33.4  &  0.5 \\

G32-49   & 2473.0 &  -26.4 &   0.1  &    +43.6  &  2.4 \\     

G173-2   & 1916.8 &  -37.6 &   0.2 \\    

G173-2   & 2474.0 &  -38.6 &   0.1 \\    

G173-2   & 2622.9 &  -35.8 &   0.1 \\    

G72-58   & 1916.8 &  +5.9  &   0.1 \\    

G72-58   & 2474.1 &  +12.6 &   0.1 \\    

G72-58   & 2623.8 &  +15.7 &   0.1 \\    

G72-59   & 1916.8 &  +15.9 &   0.2  &    +6.7   &  1.6 \\     

G173-59  & 1916.8 &  +4.3  &   0.1  &    +11.6  &  2.1 \\     

G173-59  & 2474.1 &  -1.3  &   0.1  &    +20.3  &  0.8 \\     

G173-59  & 2622.9 &  +5.9  &   0.2 \\    

G78-1    & 1916.8 &  -16.4 &   0.1  &    -5.3   &  1.4 \\     

G77-56   & 2473.1 &  +25.9 &   0.1  &    +45.3  &  1.0 \\     

G6-20    & 1916.8 &  -2.9  &   0.5  &    -35.6  &  0.7 \\     

G248-27  & 1916.8 &  -37.2 &   0.1  &    -64.2  &  1.4 \\     

G84-39   & 1916.9 &  +44.3 &   0.2  &    +92.9  &  1.8 \\     

G249-37  & 1916.9 &  +23.7 &   0.1  &    +1.9   &  1.6 \\     

G87-20   & 1916.9 &  +57.3 &   0.2  &    +72.0  &  1.3 \\     

G87-20   & 1942.9 &  +87.6 &   0.1  &    +35.2  &  0.5 \\     

G108-53  & 1942.9 &  +24.5 &   0.2  &    +33.9  &  1.3 \\     

G88-5    & 1916.9 &  -68.3 &   0.3 \\    

G88-11   & 1917.0 &  -29.5 &   0.1 \\    

G88-11   & 2032.7 &  -22.2 &   0.1 \\    

G88-11   & 2623.1 &  -28.8 &   0.1 \\    

G112-54  & 1917.0 &  +96.1 &   0.1 \\    

G112-54  & 1917.4 &  +96.5 &   0.1 \\    

G112-54  & 2032.8 &  +108.2&   0.1 \\    

G40-5    & 1917.0 &  +25.6 &   0.2 \\    

G40-5    & 2623.1 &  +33.9 &   0.1 \\    

G40-8    & 1917.0 &  -46.1 &   0.2  &    -37.1  &  0.9 \\     

G40-8    & 2032.8 &  -46.9 &   0.2  &    -37.1  &  0.5 \\     

G9-42    & 1917.1 &  +5.6  &   0.1 \\    

G9-42    & 2032.8 &  +5.9  &   0.1  &    -42.9  &  5.3 \\     

G9-42    & 2623.1 &  +6.8  &   0.1 \\    

G41-34   & 1918.0 &  +6.1  &   0.1  &    +12.5  &  2.1 \\     

G195-30  & 1918.0 &  -44.7 &   0.2 \\    

G195-30  & 1943.0 &  -69.9 &   0.2  &    +102.0 &  3.0 \\     

G195-30  & 2624.1 &  -57.4 &   0.1 \\    

G58-23   & 1917.1 &  -8.7  &   0.2  &    +16.9  &  0.8 \\     

G58-23   & 1714.8 &  -4.7  &   0.3  &    +13.1  &  0.7 \\     

G58-23   & 2032.8 &  -4.4  &   0.2  &    +11.3  &  0.4 \\     

G44-45   & 1918.0 &  +65.1 &   0.2 \\    

G44-45   & 1714.8 &  +66.8 &   0.2 \\    

G44-45   & 2032.8 &  +63.1 &   0.1 \\    

G44-45   & 2624.1 &  +58.5 &   0.1 \\    

G147-36  & 1943.0 &  +54.5 &   0.1  &    +31.9  &  1.0 \\     

G147-36  & 1915.2 &  +89.3 &   0.1  &    -30.9  &  0.9 \\     

G147-58  & 1918.1 &  -47.3 &   0.1  &    +37.7  &  1.7 \\     

G147-58  & 1943.1 &  -5.4  &   0.1 \\    

G147-58  & 1714.8 &  +25.0 &   0.2 \\    

G147-58  & 2062.8 &  -59.6 &   0.1 \\    

G59-5    & 1918.1 &  -29.1 &   0.2 \\    

G59-5    & 1714.8 &  -25.5 &   0.2 \\    

G59-5    & 2032.8 &  -27.6 &   0.1 \\    

G59-5    & 2062.8 &  -28.2 &   0.2  &    -34.4  &  0.5 \\     

G121-75  & 1714.8 &  +13.7 &   0.1  &    +31.3  &  1.2 \\     

G121-75  & 2032.8 &  +14.6 &   0.1  &    +28.5  &  0.8 \\     

G121-75  & 1918.1 &  +13.6 &   0.1  &    +29.1  &  2.6 \\     

G60-47   & 1918.1 &  +8.6  &   0.1  &    +20.3  &  1.3 \\     

G60-47   & 2032.8 &  +12.3 &   0.1  &    +14.3  &  0.6 \\     

G14-54   & 1917.1 &  -12.9 &   0.1 \\    

G14-54   & 1714.9 &  -12.9 &   0.2  &    +26.1  &  1.7 \\     

G14-54   & 2032.9 &  -5.1  &   0.1  &    -14.8  &  2.1 \\     

G62-44   & 1918.1 &  -50.2 &   0.3 \\    

G62-44   & 2032.9 &  -51.6 &   0.1  &    -45.2  &  1.9 \\     

G165-22  & 1917.1 &  -33.2 &   0.1 \\    

G165-22  & 1943.1 &  -28.7 &   0.1 \\    

G165-22  & 1714.9 &  -18.8 &   0.1 \\    

G165-22  & 2061.9 &  -20.1 &   0.1 \\    

G165-22  & 2631.2 &  -17.6 &   0.1 \\    

G62-61   & 1917.0 &  +68.3 &   0.1  &    -56.5  &  0.6 \\     

G62-61   & 1943.1 &  -0.4  &   0.1  &    +56.8  &  0.7 \\     

G62-61   & 2061.9 &  +68.3 &   0.1  &    -56.5  &  0.5 \\     

G255-45  & 1943.1 &  -43.4 &   0.1 \\    

G255-45  & 2062.8 &  -49.7 &   0.2  &    -5.2   &  2.4 \\     

G239-38  & 2061.9 &  -50.9 &   0.1 \\    

G15-6    & 1943.1 &  -62.2 &   0.1  &    -17.3  &  1.6 \\     

G15-6    & 2062.8 &  -58.6 &   0.1 \\    

G66-65   & 1943.1 &  -57.3 &   0.1 \\    

G66-65   & 2062.9 &  -59.9 &   0.1 \\    

G202-25  & 2063.0 &  -1.3  &   0.1  &    +2.6   &  1.5 \\     

G17-22   & 1715.0 &  -71.9 &   0.1 \\    

G17-22   & 2062.9 &  -25.1 &   0.1  &    -163.0 &  1.0 \\     

G227-37  & 1715.0 &  -15.3 &   0.1  &    +5.7   &  1.6 \\

G227-37  & 2472.9 &  +2.7  &   0.1  &    -28.1  &  0.4 \\     

G21-20   & 1715.0 &  -56.6 &   0.1 \\    

G21-20   & 2062.0 &  -41.9 &   0.1 \\    

G21-20   & 2473.9 &  -56.5 &   0.1 \\    

G22-7    & 1715.1 &  -74.3 &   0.1 \\    

G22-7    & 2472.9 &  -84.5 &   0.1 \\    

G262-14  & 1715.1 &  -54.4 &   0.3  &    -89.1  &  0.4 \\     

G262-14  & 2474.0 &  -71.7 &   0.2  &    -63.7  &  0.4 \\     

G230-45  & 1715.1 &  -88.0 &   0.2  &    -64.2  &  1.6 \\     

G230-45  & 2063.0 &  -88.1 &   0.2  &    -62.5  &  1.3 \\     

G262-32  & 2063.0 &  -86.4 &   0.2  &    -96.1  &  0.8 \\     

G26-38   & 2063.1 &  -11.8 &   0.1  &    -8.4   &  1.6 \\     

G26-38   & 2474.0 &  -14.3 &   0.1  &    +1.1   &  1.4 \\     

G26-38   & 2629.7 &  -16.4 &   0.1  &    +7.0   &  2.3 \\     

G18-55   & 1916.7 &  +18.7 &   0.1  &    +31.6  &  1.5 \\     

G18-55   & 1710.1 &  +19.2 &   0.2  &    +27.2  &  0.5 \\     

G67-38   & 1916.7 &  -124.7&   0.1 \\    

G67-38   & 2062.1 &  -125.6&   0.1  &    -115.3 &  1.2 \\     

G67-38   & 2474.0 &  -126.7&   0.1  &    -120.4 &  1.7 \\     

G156-75  & 1715.1 &  -45.3 &   0.1  &    -37.1  &  1.9 \\     

G156-75  & 2474.0 &  -42.5 &   0.1 \\    

G156-75  & 2629.7 &  -44.3 &   0.1 \\    

G157-21  & 2473.0 &  -39.6 &   0.1 \\    

G68-31   & 1917.7 &  -25.0 &   0.1 \\    

G68-31   & 2473.0 &  -28.8 &   0.1  &    -12.6  &  1.2 \\     

G130-10  & 1916.7 &  -61.6 &   0.1  &    -47.0  &  1.7 \\     

G130-10  & 2474.0 &  -55.7 &   0.1 \\    
\enddata
\end{deluxetable}

\clearpage
  \pagestyle{empty}
  \begin{deluxetable}{lccccc}
  \tablewidth{0pt}
  \tablecaption{Measured Mass Ratios for Systems in Table 1\label{tbl-4}}
\tablehead{
\colhead{System}&\colhead{N$_{obs}$}&\colhead{N$_{det}$}&\colhead{$q_{min}$}&
\colhead{$q_{orb}$}&\colhead{$\alpha$} }
\startdata
G130-32&   1     &  1    &$0.16\pm0.01$&$0.62\pm0.54$   &$0.12\pm0.01$\\
G32-49 &   1     &  1    &$0.29\pm0.02$&$0.36\pm0.04$   &$0.05\pm0.01$\\
G173-2 &   3     &  0    &$0.16\pm0.01$&                &          \\
G72-58 &   3     &  0    &$0.23\pm0.02$&                &         \\ 
G72-59 &   1     &  1    &$0.48\pm0.03$&$0.65\pm0.35$   &$0.28\pm0.03$ \\
G173-59&   3     &  2    &$0.62\pm0.04$&$0.62\pm0.09$   &$0.05\pm0.01$\\
G78-1  &   1     &  1    &$0.14\pm0.01$&$0.43\pm0.17$   &$0.08\pm0.01$ \\
G77-56 &   1     &  1    &$0.26\pm0.02$&$0.50\pm0.10$   &$0.13\pm0.02$ \\
G6-20  &   1     &  1    &$0.69\pm0.04$&$0.78\pm0.09$   &$0.36\pm0.01$ \\
G248-27&   1     &  1    &$0.18\pm0.01$&$0.42\pm0.09$   &$0.05\pm0.01$\\
G84-39 &   1     &  1    &$0.48\pm0.03$&$0.55\pm0.07$   &$0.16\pm0.03$ \\
G249-37&   1     &  1    &$0.32\pm0.02$&$0.43\pm0.12$   &$0.05\pm0.01$ \\
G87-20 &   2     &  2    &$0.76\pm0.06$&$0.94\pm0.06$   &$0.19\pm0.01$ \\
G108-53&   1     &  1    &$0.69\pm0.05$&$0.63\pm0.31$   &$0.10\pm0.01$ \\
G88-5  &   1     &  0    &$0.21\pm0.02$&                &              \\
G88-11 &   3     &  0    &$0.20\pm0.01$&                & \\

G112-54&   3     &  0    &$0.27\pm0.02$&                & \\
G40-5  &   2     &  0    &$0.19\pm0.02$&                & \\
G40-8  &   2     &  2    &$0.22\pm0.02$&$0.69\pm0.17$   &$0.14\pm0.01$ \\
G9-42  &   3     &  0    &$0.05\pm0.01$&                & \\
G41-34 &   1     &  1    &$0.15\pm0.02$&$0.36\pm0.28$   &$0.06\pm0.01$ \\
G195-30&   3     &  0    &$0.12\pm0.01$&                &              \\
G58-23 &   3     &  2    &$0.66\pm0.06$&$0.76\pm0.12$   &$0.16\pm0.01$ \\
G44-45 &   4     &  0    &$0.40\pm0.03$&                &              \\
G147-36&   2     &  2    &$0.59\pm0.04$&$0.57\pm0.02$   &$0.10\pm0.01$ \\
G147-58&   4     &  1    &$0.69\pm0.05$&$0.54\pm0.04$   &$0.05\pm0.01$ \\
G59-5  &   4     &  1    &$0.43\pm0.03$&$0.53\pm0.27$   &$0.27\pm0.02$ \\
G121-75&   3     &  3    &$0.37\pm0.03$&$0.29\pm0.04$   &$0.08\pm0.01$ \\
G60-47 &   2     &  2    &$0.42\pm0.03$&$0.55\pm0.23$   &$0.12\pm0.01$ \\
G14-54 &   3     &  2    &$0.19\pm0.02$&$0.20\pm0.04$   &$0.04\pm0.01$ \\
G62-44 &   2     &  0    &$0.16\pm0.01$&                &              \\
G165-22&   5     &  0    &$0.09\pm0.01$&                &              \\
G62-61 &   3     &  3    &$0.49\pm0.03$&$0.59\pm0.01$   &$0.17\pm0.01$\\
G255-45&   2     &  1    &$0.20\pm0.03$&$0.18\pm0.03$   &$0.05\pm0.01$\\
G239-38&   1     &  0    &$0.11\pm0.01$&                &          \\
G15-6  &   2     &  0    &$0.10\pm0.01$&                &          \\
G66-65 &   2     &  0    &$0.06\pm0.01$&                &              \\
G202-25&   1     &  0    &$0.33\pm0.02$&                &                \\
G227-37&   2     &  1    &$0.62\pm0.04$&$0.49\pm0.04$   &$0.27\pm0.01$\\
G21-20 &   3     &  0    &$0.13\pm0.01$&             &              \\
G22-7  &   2     &  0    &$0.21\pm0.01$&             &              \\
G262-14&   2     &  2    &$0.74\pm0.06$&$0.68\pm0.05$&$0.52\pm0.02$  \\
G230-45&   2     &  2    &$0.47\pm0.04$&$0.44\pm0.05$&$0.14\pm0.03$ \\
G262-32&   1     &  1    &$0.16\pm0.01$&$0.71\pm0.28$&$0.30\pm0.03$ \\
G26-38 &   3     &  3    &$0.24\pm0.02$&$0.30\pm0.04$&$0.06\pm0.01$  \\
G18-55 &   2     &  2    &$0.63\pm0.05$&$0.52\pm0.14$&$0.13\pm0.01$   \\
G67-38 &   3     &  2    &$0.48\pm0.03$&$0.70\pm0.20$&$0.08\pm0.01$   \\
G156-75&   3     &  0    &$0.09\pm0.01$&             &                \\
G157-21&   1     &  0    &$0.23\pm0.02$&             &             \\
G68-31 &   2     &  1    &$0.31\pm0.02$&$0.30\pm0.07$&$0.06\pm0.01$ \\
G130-10&   2     &  1    &$0.21\pm0.01$&$0.53\pm0.15$&$0.06\pm0.01$ \\
\enddata
\end{deluxetable}

\end{document}